\documentstyle[prb,aps,eqsecnum,preprint,tighten]{revtex}
\begin{document}
\draft

\title{Mean-field theory of a quasi-one-dimensional
superconductor in a high magnetic field }
\author{N. Dupuis}
\address{Laboratoire de Physique des Solides,
Universit\' e Paris-Sud, 91405 Orsay, France }
\maketitle
\begin{abstract}
At high magnetic field, the semiclassical approximation which underlies
the Ginzburg-Landau (GL) theory of the mixed state of type II
superconductors breaks down. In a quasi-1D superconductor (weakly
coupled chains system) with
an {\it open Fermi surface}, a high magnetic field stabilizes a cascade of
superconducting phases which ends in a strong reentrance of the
superconducting phase. The superconducting state evolves from a triangular
Abrikosov vortex lattice in the semiclassical regime towards a Josephson
vortex lattice in the reentrant phase.
We study the properties of these superconducting
phases from a microscopic model in the mean-field approximation. The
critical temperature is calculated in the quantum limit approximation (QLA)
where only Cooper logarithmic singularities are retained while less
divergent terms are ignored. The effects of Pauli pair breaking (PPB) and
impurity scattering are taken into account. The Gor'kov equations are solved
in the same approximation but ignoring the PPB effect.
We derive the GL expansion of the free energy and obtain the
specific heat jump at the transition. We find that each phase is first
paramagnetic and then diamagnetic for increasing field, except the reentrant
phase which is always paramagnetic. We also show that a gap opens at the
Fermi level in the quasi-particle excitation spectrum. The QLA clearly shows
how the system evolves from a quasi-2D and BCS-like behavior in the
reentrant phase towards a gapless behavior at weaker field. The calculation
is extended beyond the QLA taking into account all the pairing channels and
the validity of the QLA is discussed in detail. We show that the complete
excitation spectrum exhibits gaps at, below, and
above the Fermi level. We also calculate the current distribution.
\end{abstract}

\section{Introduction}

The equilibrium state of type II superconductors was first described by
Abrikosov using a phenomenological Ginzburg-Landau (GL) theory,
\cite{Abrikosov57} which was later justified by Gor'kov in a microscopic
model. \cite{Gorkov59} The Ginzburg-Landau-Abrikosov-Gor'kov (GLAG) theory
treats the magnetic field semiclassically and therefore can be justified
in clean materials only at high temperature or low magnetic field.
\cite{Parks69} In the last few years, there has
been a lot of work devoted to the theoretical understanding of the effect of
magnetic fields on the mean-field theory of the superconducting instability
from a completely quantum point of view.

Most of these works have been concerned with the effects of Landau level
quantization in superconductors with an isotropic dispersion law.
\cite{Rajagopal66,Tesanovic89,Tesanovic92,Dukan91,Maniv92,Ryan93,Akera91,Norman94,Rieck90,Yakovenko93} On
the one hand, the quantum effects of the field have been studied in the
vicinity of the semiclassical critical field $H_{c2}(T=0)$, with emphasis on
the precise vortex lattice structure, the quasi-particle excitation
spectrum, and the de Haas-van Alphen oscillations arising in the mixed state as
a consequence of Landau level quantization. The first observation of these
quantum magnetic oscillations in the mixed state occurred nearly twenty
years ago in the layer compound 2H-NbSe$_2$,
\cite{Graebner76} and interest has been renewed recently with their
observation in several other materials. \cite{Onuki92}
On the other hand, it has been proposed that Landau level
quantization can lead to reentrant behavior at very high magnetic field
when the cyclotron energy becomes larger than the Fermi energy ($
\omega _c\gg E_F$). \cite{Tesanovic89,Tesanovic92} This effect is
absent from the GLAG
theory which predicts a complete disappearance of the superconducting phase
due to the orbital frustration of the order parameter in the magnetic field.
This reentrant behavior originates in the suppression of the orbital
frustration when the electrons reside in only one or the few lowest Landau
levels. Indeed, when only one Landau level is occupied, the supercurrents
can be made to coincide with the orbital motion of the electrons in this
Landau level if the periodicity of the vortex lattice is approximately equal
to the orbit radius of the lowest Landau level. Moreover, in this very high
field limit, it has been argued that
the destruction of superconductivity by the Pauli pair breaking (PPB)
effect can be avoided because the effective 1D dispersion law allows one to
construct a Larkin-Ovchinnikov-Fulde-Ferrell (LOFF) \cite{Fulde64} state
which can exist far above the Pauli limited field. The reality of this
reentrant superconductivity remains however controversial
\cite{Rieck90,Yakovenko93} and there has been no experimental result up to
now.

The quantum effects of the magnetic field were also studied in the case
of quasi-one-dimensional superconductors (weakly coupled chains systems)
with {\it an open Fermi surface}.
\cite{Lebed86,Dupuis93,Dupuis94a,Dupuis94b,Dupuis94c} These effects are
especially pronounced when the zero-field critical temperature $T_{c0}$ is
smaller (but not much smaller) than the interchain coupling $t_z$ in the
direction perpendicular to the field (we will only consider this
limit in this paper). (The chains are parallel to the $x$ axis. The
external field is along the $y$ direction
and the interchain hopping $t_y$ in this direction is assumed to be much
larger than $t_z$). In this case, the superconductivity is well described
semiclassically by the anisotropic GL theory. In particular, there is no
Josephson coupling between chains even at $T=0$. Because of the quasi-1D
structure of the Fermi surface,
the semiclassical orbits in the presence of the  field
are open. Consequently, there is no Landau level quantization but the field
induces a 3D/2D crossover: \cite{Efetov83,Gorkov84,Dupuis94a} the electronic
motion remains extended along the chains and along the direction parallel
to the field, but becomes confined in the $z$ direction with an extension
$\sim ct_z/\omega _c$ ($c$ is the interchain spacing in the $z$ direction
and $\omega _c$ is the frequency of the semiclassical orbits).
This dimensional crossover is at the origin
of a very unusual phase diagram. In particular, it leads to a restoration of
time-reversal symmetry (as far as the Zeeman splitting is ignored) in
very high magnetic field ($\omega _c\gg t_z$)
which results in a reentrant behavior
of the superconducting phase with a Josephson coupling between chains.
This high-field-superconductivity can survive even in the presence of PPB
because the quasi-1D Fermi surface allows one to construct a
LOFF state (for any value of the magnetic field) which can exist far above
the Pauli limited field. Although the origin of the reentrant behavior
is very different in the quasi-1D case and in the isotropic case, in both
cases it appears as a consequence of a reduction of dimensionality, from 3D
to 2D in the quasi-1D case, and from 3D to 1D in the isotropic case.
The suppression of the orbital frustration originates in
this reduction of dimensionality.

Besides the qualitative differences between isotropic and quasi-1D
superconductors, there is also an important quantitative difference. In the
isotropic case, the temperature and magnetic ranges where quantum effects are
expected to be important are determined by the Fermi energy $E_F$. For this
reason, superconductivity is destroyed for intermediate fields, i.e. for
fields much larger than in the semiclassical regime but much smaller than in
the reentrant regime. Moreover, the
reentrant behavior can be observed only at very low temperatures and very
high fields. This restricts considerably the possible candidates to the
experimental observation of very high-field superconductivity and is one of
the reasons which explain the absence of experimental results.
In the quasi-1D case, it is the coupling $t_z$ between chains which plays
the crucial role. Since $t_z$ can be smaller than 10 K in organic conductors,
the temperature and magnetic field ranges where very high-field
superconductivity is expected can be experimentally  accessible if the
appropriate (i.e., sufficiently anisotropic) materials are chosen.
\cite{Dupuis94a,Dupuis94c} This also means that
superconductivity can survive even for intermediate fields between the GL
and the very high field regimes.
The interest in quasi-1D superconductors has been recently raised by
experimental results on the organic compound (TMTSF)$_2$ClO$_4$. \cite{Lee94}
Resistive measurements have shown an anomalous behavior of the critical field
$H_{c2}$. Although they do not give a definite answer for the
existence of high-field superconductivity in  (TMTSF)$_2$ClO$_4$,
these results might be interpreted as the signature of a high-field
superconducting phase. \cite{Dupuis94c,Lee94}

The main features of the phase diagram of a quasi-1D superconductor are now
well understood. \cite{Lebed86,Dupuis93,Dupuis94a,Dupuis94b,Dupuis94c}
Between the GL regime (where the superconducting state is a triangular
Abrikosov vortex lattice) and the reentrant phase (where the superconducting
state is a triangular Josephson vortex lattice), the magnetic field
stabilizes a cascade of superconducting
phases separated by first order transitions. In these quantum phases, the
behavior of the system (and in particular the periodicity of the order
parameter) is not determined any more by the (semiclassical) GL coherence
length $\xi _z(T)\sim 1/\sqrt{H}$ but by the transverse (i.e., perpendicular
to the chains) magnetic length $ct_z/\omega _c\sim 1/H$ ($H$ being the
external magnetic field). When entering the quantum regime,
$\omega _c\sim T$, the transverse magnetic length is much larger than the GL
coherence length $\xi _z(T)$. This results in an increase of the transverse
periodicity of the order parameter and in a strong modification of the
vortex lattice: \cite{Dupuis94a} in the quantum regime,
the amplitude of the order parameter and
the current distribution show a symmetry of a laminar type while the vortices
still describe a triangular lattice. The existence of this somehow new
superconducting state is due to the symmetry of the
one-particle wave-functions which is incompatible with the symmetry of the
Abrikosov vortex lattice. The cascade of first order phase transitions
originates in commensurability effects between the periodicity of the order
parameter (i.e., the transverse magnetic length) and the crystalline lattice
spacing.

In this paper, we study the transition line and the properties of the
superconducting phases in the mean-field approximation starting from a
microscopic model. Some of the results presented here were published
elsewhere. \cite{Dupuis94b} We assume that the superconductivity is due to an
effective attractive electron-electron interaction of the BCS type. We also
assume that the quasi-1D conductor is well described above the transition
line by the Fermi liquid theory, which justifies the use of a mean-field
theory. This situation will be realized if the system undergoes a single
particle dimensionality crossover at a temperature $T_{x^1}>T_{c0}$. Below
$T_{x^1}$, the particle-particle (Cooper) and particle-hole (Peierls)
channels decouple so that the usual mean-field (or ladder) approximation is
justified
provided that the bare parameters of the Hamiltonian are replaced by
renormalized ones in order to take into account the effects of 1D
fluctuations \cite{Bourbonnais91}
(see also Refs. \cite{Dupuis94a,Dupuis94c} for a discussion of the validity
of the mean-field approximation, in particular for the
organic conductors of the Bechgaard salts family). As pointed out by
Yakovenko, \cite{Yakovenko93} it is very important that the electrons in
the $(x,y)$ planes have a 2D behavior below $T_{x^1}$. Even if the magnetic
field suppresses the electron hopping in the $z$ direction, it has no effect
on the electron motion in the $(x,y)$ planes and the Cooper and Peierls
channels remain decoupled. \cite{Nesting} This does not exclude the
existence of
thermodynamical fluctuations, in particular in the reentrant phase ($\omega
_c\gg t_z$) where the system becomes effectively quasi-2D. In this very high
field limit, the transition from the metallic phase towards a phase with
real superconducting long-range order might be replaced by a
Kosterlitz-Thouless transition. \cite{Dupuis94a} This aspect will however
not be considered any further and we will restrict ourselves to the
mean-field analysis.

In the next section, we calculate the eigenstates and the
Green's functions of the normal phase in the presence of a uniform magnetic
field ${\bf H}(0,H,0)$. We use the gauge ${\bf A}(Hz,0,0)$
which presents the advantage to yield a very clear physical picture of the
dimensional crossover induced by the magnetic field. In Sec.\ \ref{SecTL},
we derive the transition line in the quantum limit approximation (QLA) where
only Cooper logarithmic singularities are retained while less divergent
terms are ignored. Although this approximation strongly underestimates the
critical temperature, it provides a clear physical picture of the pairing
mechanism responsible for the superconducting instability. Moreover, the
effects of disorder and PPB can easily be incorporated
in this approach. In Sec.\ \ref{SecOPQLA}, we study the superconducting
phases
in the QLA ignoring the PPB effect. We first construct a variational order
parameter using the results of Sec.\ \ref{SecTL} and then solve the Gor'kov
equations. We derive the GL expansion of the free energy and obtain the
specific heat jump at the transition. The discontinuity of the specific
heat jump at the first order transitions is related to the slope of the
first order transition lines. We find that each phase is first
paramagnetic and then diamagnetic for increasing field, except the reentrant
phase will is always paramagnetic. \cite{Melo}
We also show that a gap opens at the Fermi level in the quasi-particle
excitation spectrum. The QLA clearly shows how the system evolves from a
quasi-2D and BCS-like behavior in the reentrant phase
towards a gapless behavior at weaker field. In Sec.\ \ref{SecOPQLA}, we go
beyond the QLA. We first obtain the transition line and construct a
variational order parameter, thus recovering the results obtained in
Ref. \cite{Dupuis94a} in the gauge ${\bf A}'(0,0,-Hx)$. We then derive the GL
expansion of the free energy. We discuss the importance of the screening of
the external field by the supercurrents and also compare the results with
those obtained in the QLA. The quasi-particle excitation spectrum is
obtained from the Bogoliubov-de Gennes equations. Besides gaps which open at
the Fermi level as obtained in the QLA, gaps open below and above the Fermi
level. This excitation spectrum is very reminiscent of the
one of the field-induced-spin-density-wave (FISDW) phases which appear when
the effective electron-electron interaction is repulsive.
\cite{Yamaji85,Virosztek86,Montambaux88,Montambaux91} Finally, the
current distribution is calculated.

\section{Green's functions of the normal phase}
\label{SecGFNP}

In this section, we derive the Green's functions in the normal metallic
phase in the presence of a uniform magnetic field ${\bf H}(0,H,0)$. Contrary
to what can be found in general in the literature concerning quasi-1D
conductors in a magnetic field, we work in the gauge ${\bf A}(Hz,0,0)$.
The one-particle Hamiltonian is obtained from
the Peierls substitution ${\cal H}_0=E({\bf k} \rightarrow -i{\bf
\nabla }-e{\bf A})$. The dispersion law is given by ($\hbar =k_B=1$
throughout the paper and the Fermi energy $E_F$ is chosen as the origin of the
energies)
\begin{equation}
E({\bf k})=v(\vert k_x \vert -k_F)+t_y\cos (k_yb) +t_z\cos (k_zc) \,,
\label{disp}
\end{equation}
where $v$ is the Fermi velocity for the motion along the chains ($x$ axis)
and $t_y$, $t_z$ are the couplings between chains separated by the distance
$b$ and $c$. The condition $t_y,t_z\ll E_F$ ensures that the Fermi surface
is open. Except in a few cases which will be pointed out when necessary,
we will not explicitly consider the $y$
direction parallel to the magnetic field which does not play any role for a
linearized dispersion law (as long as Cooper pairs are formed with states of
opposite momenta in this direction). In order to take into account the $y$
direction, we just have to replace the 2D density of states per spin
$N(0)=1/\pi vc$ by its 3D value $1/\pi vbc$. It should be noted here that no
generality is lost at the mean-field level when studying a 2D system instead
of a 3D system. This is due to the fact that the kinetic energy mainly comes
from the motion along the chains, which is not affected by the field (see
below), so that the electron-electron interaction can still be treated
in perturbation for a 2D system. This should be contrasted with 2D isotropic
systems in high magnetic field where the perturbative treatment (i.e. the
mean-field analysis) of the superconducting instability is highly
questionable. \cite{Tesanovic92,Yakovenko93}

Since the magnetic field does not couple the two sheets
of the Fermi surface, \cite{Dupuis92} we can write an Hamiltonian for each
sheet of the Fermi surface:
\begin{equation}
{\cal H}_{0,\sigma }^\alpha =v(-i\alpha \partial _x -k_F)+\alpha \hat m
\omega _c +t_z\cos (-ic\partial _z) +\sigma h  \,,
\label{Hami}
\end{equation}
where $\alpha =+ \,(-)$ labels the right (left) sheet of the Fermi surface.
$\hat m$ is the (discrete) position operator in the $z$ direction and
$\sigma =+\,(-)$ for $\uparrow (\downarrow )$ spin. $\sigma h=\sigma \mu
_BH$ is the Zeeman energy for a $g$ factor equal to 2. We have introduced
the energy $\omega _c=Gv$ where $G=-eHc$ is a magnetic wave vector.
The operator $-i\partial _x$ commutes with the Hamiltonian so that the
momentum $k_x$ along the $x$ axis is a good quantum number. We therefore
look for a solution $\phi _{k_x}^\alpha (x,m)=e^{ik_xx}\phi
_{k_x}^\alpha (m)$. The
Fourier transform $\phi _{k_x}^\alpha (k_z)$ of $\phi _{k_x}^\alpha (m)$ is
solution of the Schr\"odinger equation (setting $c=1$ for simplicity)
\begin{equation}
\lbrack i\alpha \omega _c\partial _{k_z}+t_z \cos (k_z) \rbrack
\phi _{k_x}^\alpha (k_z)=\tilde \epsilon \phi _{k_x}^\alpha
(k_z)  \,,
\label{EQS1}
\end{equation}
where $\tilde \epsilon =\epsilon -v(\alpha
k_x-k_F)-\sigma h$ and $\epsilon $ is the eigenenergy of
the eigenstate $\phi _{k_x}^\alpha (x,m)$. The solution of (\ref{EQS1}) is
\begin{equation}
\phi _{k_x}^\alpha (k_z) \sim e^{-i {\alpha \over {\omega _c}}
\lbrack \tilde \epsilon k_z -t_z\sin (k_z) \rbrack  }
\end{equation}
up to a normalization factor. Going back to real space, we obtain
\begin{equation}
\phi _{k_x}^\alpha (x,m) \sim \int _{-\pi }^\pi {{dk_z} \over {2\pi }}
e^{ ik_z(m- {\alpha \over {\omega _c}}\tilde \epsilon )
+i {\alpha \over {\omega _c}}t_z \sin (k_z) } \,.
\end{equation}
$\phi _{k_x}^\alpha (x,m)$ is non zero only if $\tilde \epsilon =\alpha l
\omega _c$ where $l$ is an integer. Therefore, the normalized eigenstates
and the eigenenergies of the Hamiltonian ${\cal H}_{0,\sigma }^\alpha $ are
given by (writing explicitly the interchain spacing $c$)
\begin{eqnarray}
\phi _{k_x,l}^\alpha ({\bf r})&=&{1 \over \sqrt{cL_x}} e^{ik_x x}
J_{l-m}(\alpha \tilde t) \label{Estates} \,, \\
\epsilon _{k_x,l,\sigma }^\alpha &=&v(\alpha k_x-k_F)+\alpha l\omega _c
+\sigma h  \,,
\label{spectre}
\end{eqnarray}
where ${\bf r}=(x,m)$ and $J_l$ is the $l$th
order Bessel function. $L_x$ is the length of the system in the $x$
direction and $\tilde t=t_z/\omega _c$ is a reduced interchain coupling.
The state $\phi _{k_x,l}^\alpha $ is localized around the
$l$th chain with a spatial extension in the $z$ direction of the order of
$\tilde t c$ which corresponds to the amplitude of the semiclassical orbits.
\cite{Dupuis94a} The advantage of working with the vector potential ${\bf
A}(Hz,0,0)$ is now clear: in this gauge, the eigenstates are localized,
which corresponds to the real physical behavior of the particles as can be
seen
by examining the one-particle Green's function in real space.\cite{Dupuis92}
Moreover, since the momentum $k_x$ remains a good quantum number, the
motion along the chains is in some sense trivial. Note that the states $\phi
_{k_x,l}^\alpha $ can be deduced by the appropriate gauge
transformation from the localized
states introduced by Yakovenko \cite{Yakovenko87} in the gauge ${\bf
A}'(0,0,-Hx)$.

In the Hamiltonian (\ref{Hami}), the magnetic field appears only through the
additional term $\alpha \hat m \omega _c$. The effect of the magnetic field
is therefore similar to an ``electric field'' $-\alpha vH$ whose sign
would depend on the sheet of the Fermi surface. This fictitious electric
field
introduces an additional ``potential energy" $\alpha m\omega _c$ on each
chain $m$ which competes with the hopping term $t_z\cos (-ic\partial _z)$
and tends to prevent the electronic motion in the $z$ direction.
The localization in the $z$ direction is almost complete when the
difference of ``potential energy" $\omega _c$
between two chains is much larger than the transfer integral
$t_z$ between chains ($\omega _c \gg t_z$). The fact that an electric field
can localize the wave functions (as obtained here in a tight-binding model)
has been known for a long time. \cite{Wannier60} This
effect has recently attracted a lot of attention in connection with the
studies of semiconductor superlattices submitted to
an electric field. \cite{Mendez88} The quantized spectrum which
results from this localization is known as a Wannier-Stark ladder. The
semiclassical picture of this effect yields the well-known Bloch
oscillations of a band electron in an electric field. Continuing this
analogy
between electric field and magnetic field in a quasi-1D conductor, we can
interpret the semiclassical trajectories $z=z_0 +c(t_z/\omega
_c)\cos (Gx)$ obtained from ${\cal H}_0$ as Bloch
oscillations of the electrons in the magnetic field. Most of the effects
which are induced by the magnetic field in a quasi-1D conductor can be
understood as the consequence of these Bloch oscillations.

In the next section, it will be useful to use Green's functions in the
representation of the states  $\phi _{k_x,l}^\alpha $. Introducing
the creation and annihilation operators $b_{k_x,l,\sigma }^{\alpha \dagger
}$, $b_{k_x,l,\sigma }^\alpha $ of a particle of spin $\sigma $ in the
state $\phi _{k_x,l}^\alpha $, we define the Matsubara Green's function
$G_\sigma ^\alpha (k_x,l,\omega )$ by
\begin{eqnarray}
G_\sigma ^\alpha (k_x,l,\omega ) &=& - \int _0^{1/T} d\tau \, e^{i\omega
\tau
} \langle T_\tau b_{k_x,l,\sigma }^\alpha (\tau )b_{k_x,l,\sigma }^{\alpha
\dagger }(0) \rangle  \nonumber \\
&=& (i\omega - \epsilon ^\alpha _{k_x,l,\sigma })^{-1} \,,
\label{GF1}
\end{eqnarray}
where $\omega \equiv \omega _n=\pi T(2n+1)$ is a Matsubara frequency.

\section{Transition line in the quantum limit approximation }
\label{SecTL}

The exact mean-field critical temperature $T_c$ has been calculated
numerically and discussed in detail
elsewhere. \cite{Dupuis93,Dupuis94a,Dupuis94c} In this section,
we calculate $T_c$ in the QLA, where only Cooper logarithmic
singularities are retained while less divergent terms are ignored. Although
this approximation yields a critical temperature several orders of magnitude
smaller than the exact mean-field result, it provides a clear physical
picture of the pairing mechanism responsible for the superconducting
instability. It will also allow us to give a simple description
of the properties of the ordered phase below $T_c$ (see Sec.\
\ref{SecOPQLA}).
Moreover, the effect of disorder can be easily incorporated at this level of
approximation.

The total Hamiltonian is now ${\cal H}_0+{\cal H}_{\rm int}$ where the
effective attractive electron-electron Hamiltonian is described by the BCS
model with coupling constant $\lambda >0$:
\begin{equation}
{\cal H}_{\rm int}=-{\lambda \over 2} \sum _{\alpha ,\alpha '=\pm ,\sigma }
\int d^2{\bf r} \,
\psi _\sigma ^{\alpha '\dagger }({\bf r})
\psi _{\overline \sigma }^{\overline \alpha '\dagger }({\bf r})
\psi _{\overline \sigma }^{\overline \alpha }({\bf r})
\psi _\sigma ^\alpha ({\bf r}) \,.
\end{equation}
We use the notation $\int d^2{\bf r} =c\sum _m \int dx$ and $\overline
\alpha =-\alpha $, $\overline \sigma =-\sigma $.
The $\psi _\sigma ^\alpha ({\bf r})$'s are fermionic operators for
particles moving on the sheet $\alpha $ of the Fermi surface. The
interaction is effective only between particles whose energies are within
$\Omega $ of the Fermi level.

We first note that the superconducting instability can be qualitatively
understood from the Wannier-Stark ladder (\ref{spectre}).
In zero-field, time-reversal
symmetry ensures that $E_{\uparrow }({\bf k})=E_{\downarrow }(-{\bf k})$ so
that the pairing at zero total momentum presents the usual (Cooper)
logarithmic singularity $\sim \ln (2\gamma \Omega /\pi T)$ ($\gamma
\sim 1.781$ is the exponential of the Euler constant)
which results in an instability of the metallic
state at a finite temperature $T_{c0}$. A finite magnetic field breaks down
time-reversal symmetry. Nonetheless, we still have $\epsilon
_{k_x,l_1,\uparrow }^\alpha =\epsilon _{q_x-k_x,l_2,\downarrow
}^{\overline \alpha }$ for $q_x=-(l_1+l_2)G$ (if we ignore the Zeeman
splitting). Thus, whatever the value of the field, some pairing
channels present the Cooper singularity
if the total momentum along the chain $q_x$ is a
multiple of $G$. This results in logarithmic divergences at low temperature
in the linearized gap equation, which destabilize the metallic state
at a temperature $T_c$ ($0<T_c<T_{c0}$). \cite{Lebed86,Dupuis93,Dupuis94a}
This reasoning holds also in presence of the Zeeman splitting if we consider
pairing at total momentum $q_x=-(l_1+l_2)G \pm 2h/v$ for two bars $l_1$,
$l_2$ of the Wannier-Stark ladder. The shift $\pm 2h/v$ of the total pairing
momentum, which displaces the Fermi surfaces of spin $\uparrow $ and spin
$\downarrow $ relative to each other, partially compensates the PPB effect
and yields to the formation of a LOFF state.
\cite{Dupuis93,Dupuis94a,Dupuis94c}

Besides the most singular channels which present the Cooper singularity,
there exist less singular channels with singularities $\sim \ln \vert
2\Omega
/n\omega _c \vert $ ($n\ne 0$) for $T\ll \omega _c$ as will be shown
below. In this quantum limit ($\omega _c \gg T$),
a natural approximation consists in retaining only the most singular
channels. This QLA has been used previously in the mean-field theory
of isotropic superconductors in a high magnetic field. \cite{Tesanovic92}

It is worth pointing out that the same kind of reasoning can explain the
appearance of the FISDW phases in the presence of
a repulsive electron-electron interaction. \cite{Montambaux91}
Even if we add to the Hamiltonian
a second neighbor hopping term $t_z'\cos (2k_zc)$ (assumed to be large enough
to suppress any SDW instability in zero magnetic field), the spectrum
remains a Wannier-Stark ladder (Eq.\ (\ref{spectre})) although the expression
of the eigenstates differ from (\ref{Estates}). Since $\epsilon ^-
_{k_x,l_1,\sigma }=-\epsilon ^+ _{Q_x+k_x,l_2,\overline \sigma
}$ for $Q_x=2k_F+(l_1-l_2)G$, some electron-hole pairing channels
present logarithmic singularities if $Q_x-2k_F$ is a multiple of $G$.
At low temperature and high enough field,
this leads to an instability of the metallic phase with respect to a SDW
phase at wave vector $Q_x=2k_F+NG$ ($N$ integer). Thus, the gauge ${\bf
A}(Hz,0,0)$ provides a very natural picture of the quantized nesting
mechanism \cite{Heritier84} which is at the origin of the FISDW phases in
quasi-1D conductors.
\cite{Montambaux87} The QLA approximation has also been used in this
context where it is known as the single gap approximation (SGA).
\cite{Virosztek86,Montambaux88,Maki86}

\subsection{Without disorder and without PPB }
\label{SubSecWDWPPB}

We first consider the simplest case where both PPB and disorder are
neglected.
In order to obtain the critical temperature, we consider the
two-particle vertex function $\Gamma ^{\alpha \alpha '}({\bf r}_1,{\bf
r}_2;{\bf r}'_1,{\bf r}'_2)$ for a pair of particles on opposite sides of
the Fermi surface and with opposite spins and Matsubara frequencies.
$\Gamma ^{\alpha \alpha '}$ is evaluated in the ladder approximation shown
diagrammatically in Fig.\ \ref{FigLA}. We first write the two-particle
vertex function in the representation of the eigenstates
$\phi _{k_x,l}^\alpha $:
\begin{eqnarray}
\Gamma ^{\alpha \alpha '}({\bf r}_1,{\bf r}_2;{\bf r}'_1,{\bf r}'_2) &=&
{1 \over {cL_x}} \sum _{q_x} \sum _{k_x,l_1,l_2} \sum _{k'_x,l'_1,l'_2}
\Gamma ^{\alpha \alpha '}_{q_x}(l_1,l_2;l'_1,l'_2)
\nonumber \\ & & \times
\phi _{k_x,l_1}^\alpha ({\bf r}_1)
\phi _{q_x-k_x,l_2}^{\overline \alpha }({\bf r}_2)
\phi _{k'_x,l'_1}^{\alpha '}({\bf r}'_1)^*
\phi _{q_x-k'_x,l'_2}^{\overline \alpha '}({\bf r}'_2)^*  \,.
\end{eqnarray}
Here we have used the fact that the total momentum $q_x$ of the Cooper
pair along the chains is conserved. In the ladder approximation, the vertex
$\Gamma ^{\alpha \alpha '}_{q_x}(l_1,l_2;l'_1,l'_2)$ is solution of the
equation
\begin{eqnarray}
\Gamma ^{\alpha \alpha '}_{q_x}(l_1,l_2;l'_1,l'_2)  &=&
\langle l_1,\alpha ;l_2,\overline \alpha \vert {\cal H}_{\rm int} \vert
l'_1,\alpha ';l'_2,\overline \alpha ' \rangle
\nonumber \\ & &
- \sum _{\alpha '',l''_1,l''_2}
\langle l_1,\alpha ;l_2,\overline \alpha \vert {\cal H}_{\rm int} \vert
l''_1,\alpha '';l''_2,\overline \alpha '' \rangle
\nonumber \\ & & \times
\chi ^{\alpha ''}(q_x+(l''_1+l''_2)G)
\Gamma ^{\alpha ''\alpha '}_{q_x}(l''_1,l''_2;l'_1,l'_2)  \,,
\label{EQVertex1}
\end{eqnarray}
where
\begin{equation}
\chi ^{\alpha ''}(q_x+(l''_1+l''_2)G)={T \over {L_xc}} \sum _{\omega ,k''_x}
G_\uparrow ^{\alpha ''}(k''_x,l''_1,\omega )
G_\downarrow ^{\overline \alpha ''}(q_x-k''_x,l''_2,-\omega )
\end{equation}
is the two-particle propagator in the representation of the states $\phi
_{k_x,l}^\alpha $. Using the expression (\ref{GF1}) for the
Green's function $G_\sigma ^\alpha (k_x,l,\omega )$, we have
\begin{equation}
\chi ^\alpha (q_x)= {{N(0)} \over 2} \Biggl \lbrack \ln
\Biggl ( {{2\gamma \Omega } \over {\pi T}} \Biggr ) +
\Psi \Biggl ( {1 \over 2} \Biggr ) - {\rm Re} \,
\Psi \Biggl ( {1 \over 2}+ {{\alpha vq_x} \over {4i\pi T}} \Biggr )
\Biggr \rbrack \,,
\label{PP0}
\end{equation}
where $\Psi $ is the digamma function.
$N(0)=1/\pi vc$ is the density of states per spin at the Fermi level.
Note that $\chi (q_x)=\sum _\alpha
\chi ^\alpha (q_x)$ is the pair susceptibility
at zero magnetic field evaluated at the total momentum $q_x$.
The first term on the right-hand side of (\ref{EQVertex1})
is the two-particle matrix element of the electron-electron interaction
${\cal H}_{\rm int}$:
\begin{eqnarray}
& \langle l_1,\alpha ;l_2,\overline \alpha \vert {\cal H}_{\rm int} \vert
l'_1,\alpha ';l'_2,\overline \alpha ' \rangle
\delta _{k_{1x}+k_{2x},k'_{1x}+k'_{2x}} & \nonumber \\
=&  -\lambda \int d^2 {\bf r} \,
\phi ^\alpha _{k_{1x},l_1} ({\bf r})^*
\phi ^{\overline \alpha }_{k_{2x},l_2} ({\bf r})^*
\phi ^{\alpha '}_{k'_{1x},l'_1} ({\bf r})
\phi ^{\overline \alpha '}_{k'_{2x},l'_2} ({\bf r}) &
\nonumber \\
=&-\lambda \delta _{k_{1x}+k_{2x},k'_{1x}+k'_{2x}}
\alpha ^{l_1-l_2} {\alpha '}^{l'_1-l'_2}
\int _0^{2\pi } {{dx} \over {2\pi }}
e^{i(l_1+l_2-l'_1-l'_2)x }
J_{l_1-l_2}(2\tilde t \cos x)
J_{l'_1-l'_2}(2\tilde t \cos x)    \,. &
\end{eqnarray}

It is clear from (\ref{EQVertex1}) that Cooper logarithmic singularities
$\chi ^\alpha (0) \sim \ln (2\gamma \Omega /\pi T)$ appear when the
intermediate two-particle state corresponds to
$q_x+2L''G=0$ (which imposes $q_x$ to be a multiple of $G$) where $L''$ is
the center of gravity of the Cooper pair with respect
to the $z$ direction. Intermediate states with $q_x+2L''G=nG$ ($n \neq 0$)
lead to weaker logarithmic singularities $\chi ^\alpha (nG)\sim \ln
\vert 2\Omega /n\omega _c \vert $ for $\omega _c \gg T$. The
QLA, which is expected to be valid for $\omega _c \gg T$, restricts the
Hilbert space to the intermediate states such that $q_x+2L''G=0$.
Since $q_x$ is a constant of motion, the
center of gravity $L=(l_1+l_2)/2=-q_x/2G$ also becomes a constant of motion
of the Cooper pair in the QLA. Thus, (\ref{EQVertex1})
reduces to
\begin{equation}
\Gamma _{q_x(L),L}^{\alpha \alpha '}(l,l') = -\lambda
V_{l,l'}^{\alpha \alpha '}
+{\lambda \over 2} \chi (0) \sum _{\alpha '',l''}
V_{l,l''}^{\alpha \alpha ''}
\Gamma _{q_x(L),L}^{\alpha ''\alpha '}(l'',l')  \,,
\end{equation}
where $l=l_1-l_2$ and $l'=l'_1-l'_2$ describe the relative motion of the
pair in the $z$ direction. $q_x(L)=-2LG$
and $V_{l,l'}^{\alpha \alpha '}=\alpha ^l {\alpha
'}^{l'}V_{l,l'}$ with
\begin{equation}
V_{l,l'} =
\int _0^{2\pi } {{dx} \over {2\pi }}
J_l(2\tilde t\cos x)
J_{l'}(2\tilde t\cos x)   \,.
\label{MatrixV}
\end{equation}
Note that $V_{l,l'}^{\alpha \alpha '}$ is independent of the center of
gravity $L$ of the Cooper pairs. To eliminate the dependence on $\alpha $
and $\alpha '$, we write $\Gamma _{q_x(L),L}^{\alpha \alpha
'}(l,l')=-\lambda \alpha ^l {\alpha '}^{l'} \Gamma _{q_x(L),L}(l,l')$. $\Gamma
_{q_x(L),L}$ is solution of the equation
\begin{equation}
\Gamma _{q_x(L),L}(l,l')=V_{l,l'} + \lambda \chi (0) \sum _{l''}
V_{l,l''} \Gamma _{q_x(L),L}(l'',l') \,.
\end{equation}
The preceding matrix equation is solved
by introducing the orthogonal transformation $U_{l,l'}$ which
diagonalizes the matrix $V_{l,l'}$: $(U^{-1}VU)_{l,l'}=\delta _{l,l'}\bar
V_{l,l}$. The matrix $\bar \Gamma _{q_x(L),L}=U^{-1}\Gamma _{q_x(L),L}U$
is diagonal:
\begin{equation}
\bar \Gamma _{q_x(L),L}(l,l')=\delta _{l,l'} {{\bar V _{l,l}} \over
{1-\lambda \chi (0) \bar V_{l,l}}} \,,
\end{equation}
and we obtain
\begin{equation}
\Gamma _{q_x(L),L}^{\alpha \alpha '}(l,l')=-\lambda
\alpha ^l {\alpha '}^{l'} \sum _{l''}
{{U_{l,l''} \bar V_{l'',l''} (U^{-1})_{l'',l'} }
\over {1-\lambda \chi (0) \bar V_{l'',l''} }}  \,.
\label{vertex}
\end{equation}
The metallic state becomes
instable when a pole appears in the two-particle vertex function. Using
$\chi
(0)=N(0)\ln (2\gamma \Omega /\pi T)$, we obtain the critical temperature
\begin{equation}
T_c={{2\gamma \Omega } \over {\pi }} \exp \Biggl ( {-1 \over {\lambda N(0)
\bar V_{l_0,l_0}}} \Biggr ) \,,
\label{TCQLA}
\end{equation}
where $\bar V_{l_0,l_0}$ is the highest eigenvalue of the matrix $V$.
The critical temperatures are shown in Fig.\ \ref{FigTCQLA} for the two
highest eigenvalues of $V$. The parameters used in the numerical
calculations ($T_{c0}=1.5$ K and $t_z=20$ K) are the same as those of
Fig.\ 1 and Figs.\ 4-10 of Ref.\ \cite{Dupuis94a}. Fig.\ \ref{FigTCQLA}
clearly indicates that there are two lines of
instability competing with each other and leading to a cascade of first
order transitions in agreement with the exact mean-field calculation of
$T_c$. \cite{Dupuis94a} The existence of two lines of instability results
from the fact that $V_{l,l'}=0$ if $l$ and $l'$ do not have the same parity.
Diagonalizing the matrix $V_{l,l'}$ is then equivalent to separately
diagonalizing the matrices $V_{2l,2l'}$ and $V_{2l+1,2l'+1}$. In the
following,
we label these two lines by $l_0=0,1$ so that $\bar V_{l_0,l_0}= \max _l
\bar V_{2l+l_0,2l+l_0}$. Since $2L=l_1+l_2$ and $l_1-l_2$ have the same
parity,
$L$ is integer (half-integer) for $l_0=0$ ($l_0=1$) and can be written as
$L=-l_0/2+p$ with $p$ integer. Correspondingly, we have $q_x(L)=(l_0-2p)G$.
It is clear that the instability line $l_0$ corresponds to the instability
line $Q=l_0G$ which was previously obtained in another approach where the
magnetic Bloch wave vector $Q$ plays the role of a pseudo-momentum
for the Cooper pairs in the magnetic field. \cite{Dupuis94a}

As can be seen from Fig.\ \ref{FigTCQLA}, $T_c$ calculated in the QLA is
several orders of magnitude below the exact critical temperature except in
the reentrant phase: it has been pointed out previously that in general the
QLA strongly underestimates the critical temperature. \cite{Montambaux88}
In the QLA, we neglect intermediate pair states with a center of gravity
$L''\neq L=-q_x(L)/2G$. Since the one-particle states $\phi ^\alpha
_{k_x,l}$ are localized, $\vert
L''-L\vert $ is bounded by $\sim t_z/\omega _c$. This means that the QLA
neglects the logarithmic divergences $\ln (2\Omega /\omega _c)$,
$\ln (\Omega /\omega
_c)$, ..., $\ln (2\Omega /t_z)$. There are therefore two cases where the QLA
becomes quantitatively correct. Either $\omega _c> t_z$ (which
corresponds to the reentrant phase) so that the
only logarithmic divergence to be considered is the Cooper singularity $\ln
(2\gamma \Omega /\pi T)$. Or the cutoff energy $\Omega $ is sufficiently low
for the condition $\omega _c \sim \Omega $ to hold. In a conventional
(isotropic) superconductor where the attractive electron-electron
interaction is due to the electron-phonon coupling, $\Omega $ is the Debye
frequency and the condition $\omega _c \sim \Omega $ can never be satisfied for
reasonable values of the magnetic field. In a quasi-1D superconductor, it
has been argued that $\Omega \sim T_{x^1}$, where $T_{x^1}$ is the single
particle dimensionality crossover temperature below which the system becomes
3D. \cite{Bourbonnais91} The reason is that the superconducting
instability cannot develop at energies $\epsilon >T_{x^1}$ (or equivalently
at length scales $<v/T_{x^1}$) where the behavior of the system is 1D.
In organic superconductors like the Bechgaard salts, $T_{x^1}$ can be of the
order of $10-30$ K, so that the condition $\omega _c\sim \Omega $ could be
realized in particular cases although this remains quite unlikely. Moreover,
in the weak coupling limit $T_c\ll \Omega $, the QLA can never be
quantitatively correct when entering the quantum regime ($\omega _c\sim T$).

{}From (\ref{vertex}), one can see that the superconducting condensation
in the channel $q_x(L),L,l_0$ corresponds to the following spatial
dependence for the order parameter
\begin{equation}
\Delta _{q_x(L),L,l_0}({\bf r})\sim \sum _{\alpha ,l} \alpha ^l U_{l,l_0}
\phi _{k_x,L+{l \over 2}}^\alpha ({\bf r})
\phi _{q_x(L)-k_x,L-{l \over 2}}^{\overline \alpha }({\bf r})  \,.
\label{po1}
\end{equation}
Noting that the matrices $V$ and $U$ have a range of the order of $\tilde t$
(i.e., $V_{l,l'}$, $U_{l,l'}$ are important for $\vert l \vert, \, \vert l'
\vert < \tilde t$), one can see that $\Delta _{q_x(L),L,l_0}$ has the form
of a strip extended in
the direction of the chains and localized in the perpendicular direction on
a length of the order of $c\tilde t$. This is not surprising since
$\Delta _{q_x(L),L,l_0}$ results from pairing between the localized states
$\phi _{k_x,l}^\alpha $. The expression (\ref{po1}) of the order parameter
at $T_c$ will be used in Sec.\ \ref{SecOPQLA} to construct a variational
order parameter describing the ordered phase below $T_c$.

\subsection{Effect of disorder}
\label{SubSecEOD}

We evaluate the effect of disorder on the critical temperature
calculated in the QLA. In presence of impurity scattering, the pair
propagator appearing in the integral equation for the vertex function
$\Gamma
^{\alpha \alpha '}$ has to be modified by self-energy and vertex corrections
\cite{Gorkov60} as shown diagrammatically in Fig.\ \ref{FigPP}. In the Born
approximation, the self-energy is given by \cite{Dupuis92}
\begin{equation}
\Sigma ^{\rm dis}(\omega )=-{i \over {2 \tau }} {\rm sgn}(\omega ) \,,
\end{equation}
where $1/\tau =2\pi N(0)n_iV_i^2$. $n_i$ is the impurity density and
$V_i$ is the strength of the electron-impurity interaction which is
assumed to be local in real space. The elastic
scattering time $\tau $ is not affected by the magnetic field because the
density of states per spin $N(\epsilon ,H)=N(0)$ is magnetic field
independent. Thus,
self-energy corrections can be taken into account by the usual replacement
$\omega \rightarrow \tilde \omega =\omega +{\rm sgn}(\omega )/2\tau $ in the
expression of the one-particle Green's function. Because of the vertex
corrections shown in Fig.\ \ref{FigPP}, the pair propagator $\Pi _{\tilde
\omega }^{\alpha _1 \alpha _2}(l_1,l_2)$ in the pairing channel $q_x(L),L$
is determined by the matrix equation
\begin{equation}
\Pi _{\tilde \omega }^{\alpha _1 \alpha _2}(l_1,l_2) =
\delta _{\alpha _1,\alpha _2} \chi _{\tilde \omega }^{\alpha _1}(0)
+u_0 \chi _{\tilde \omega }^{\alpha _1}(0)
\sum _{\alpha _3,l_3} V_{l_1,l_3}^{\alpha _1\alpha _3}
\Pi _{\tilde \omega }^{\alpha _3 \alpha _2}(l_3,l_2)  \,,
\label{ME1}
\end{equation}
where $u_0=1/2\pi N(0)\tau $. The variables $l_i$ refer to the relative
motion of the Cooper pair in the $z$ direction. In writing (\ref{ME1}), we
have used the fact that in the QLA, the only states which are allowed
satisfy $q_x(L)=-2LG$ where $L$ is the center of gravity of the pair with
respect to the $z$ direction. $\chi _{\tilde \omega }^\alpha (0)$ is defined by
\begin{equation}
\chi _{\tilde \omega }^\alpha (0)=
{1 \over {cL_x}} \sum _{k_x}
G_\uparrow ^\alpha (k_x,L+{{l_1} \over 2},\tilde \omega )
G_\downarrow ^{\overline \alpha }(q_x(L)-k_x,L-{{l_1} \over 2},-\tilde
\omega ) \,.
\end{equation}
Introducing the matrix
\begin{equation}
\Pi _{\tilde \omega }(l_1,l_2)=
\sum _{\alpha _1,\alpha _2} (\alpha _1\alpha _2)^{l_1} \Pi _{\tilde \omega
}^{\alpha _1 \alpha _2}(l_1,l_2) \,,
\end{equation}
the matrix equation (\ref{ME1}) becomes
\begin{equation}
\Pi _{\tilde \omega }(l_1,l_2)=
\chi _{\tilde \omega }(0)
+u_0 \chi _{\tilde \omega }(0) \sum _{l_3} V_{l_1,l_3}
\Pi _{\tilde \omega }(l_3,l_2) \,,
\end{equation}
where $\chi _{\tilde \omega }(0)=\sum _\alpha \chi _{\tilde \omega }^\alpha
(0)$. In order to obtain the preceding equation, we used the property that
$V_{l,l'}$ is non zero only if $l$ and $l'$ have the same parity. The matrix
$\Pi _{\tilde \omega }$ is diagonalized by the transformation
$U$:
\begin{eqnarray}
\bar \Pi _{\tilde \omega }(l_1,l_2) &=&
(U^{-1}\Pi _{\tilde \omega }U)_{l_1,l_2} \nonumber \\
&=& \delta _{l_1,l_2} {{\chi _{\tilde \omega }(0)} \over { 1-u_0\chi
_{\tilde \omega }(0) \bar V_{l_1,l_1} }} \,.
\label{PP1}
\end{eqnarray}

In presence of impurity scattering, the two-particle vertex function is
determined in the QLA by the equation
\begin{equation}
\Gamma _{q_x(L),L}^{\alpha _1\alpha _2}(l_1,l_2) =
-\lambda V_{l_1,l_2}^{\alpha _1\alpha _2}
+\lambda T \sum _\omega \sum _{\alpha _3,l_3} \sum _{\alpha _4,l_4}
V_{l_1,l_3}^{\alpha _1\alpha _3}
\Pi _{\tilde \omega }^{\alpha _3\alpha _4}(l_3,l_4)
\Gamma _{q_x(L),L}^{\alpha _4\alpha _2}(l_4,l_2)  \,.
\end{equation}
The dependence on the indices $\alpha _i$ is suppressed by writing $\Gamma
^{\alpha _1\alpha _2}(l_1,l_2)=-\lambda \alpha _1^{l_1}\alpha _2^{l_2}
\Gamma (l_1,l_2)$. Using the property that $\Pi _{\tilde \omega }^{\alpha
\alpha '}(l,l')$ is non zero only
if $l$ and $l'$ have the same parity, we obtain
\begin{equation}
\Gamma _{q_x(L),L}(l_1,l_2)=V_{l_1,l_2} + \lambda T \sum _\omega
\sum _{l_3,l_4} V_{l_1,l_3} \Pi _{\tilde \omega }(l_3,l_4)
\Gamma _{q_x(L),L}(l_4,l_2) \,.
\end{equation}
The preceding matrix equation is diagonalized by the transformation
$U$:
\begin{eqnarray}
\bar \Gamma _{q_x(L),L}(l,l')&=&(U^{-1}\Gamma _{q_x(L),L}U)_{l,l'}
\nonumber \\ &=& \delta _{l,l'} {{\bar V _{l,l}} \over
{1-\lambda \bar V_{l,l} T \sum _\omega \bar \Pi _{\tilde \omega }(l,l) }}
\label{Vertex1}
\end{eqnarray}
{}From (\ref{PP1}) and (\ref{Vertex1}), we see that the appearance of a pole
in the two-particle vertex function corresponds to
\begin{equation}
1-\lambda \bar V_{l,l} T \sum _\omega
{{\chi _{\tilde \omega }(0)} \over
{1-u_0\bar V_{l,l} \chi _{\tilde \omega }(0) }} =0 \,.
\label{pole}
\end{equation}
This equation determines the critical temperature in the pairing channel
$q_x(L),L,l$. As in the pure system, the highest $T_c$ is obtained for
$l=l_0$ defined by $\bar V_{l_0,l_0}={\rm max}_l \bar V_{l,l}$.
The result (\ref{pole}) can be expressed  as
\begin{equation}
N(0) \ln \Biggl ( {{T_c^{\rm dis}} \over {T_c}} \Biggr ) =
T_c^{\rm dis} \sum _\omega  {{\chi _{\tilde \omega }(0)} \over
{1-u_0\bar V_{l_0,l_0} \chi _{\tilde \omega }(0) }}
-T_c^{\rm dis} \sum _\omega  \chi _\omega (0) \,,
\end{equation}
where $T_c^{\rm dis}$ is the critical temperature in presence of disorder
and $T_c$ the critical temperature of the pure system given by (\ref{TCQLA}).
Using
\begin{equation}
\chi _\omega (0) = {{\pi N(0)} \over {\vert \omega \vert}} \,,
\end{equation}
we obtain
\begin{equation}
\ln \Biggl ( {{T_c^{\rm dis}} \over {T_c}} \Biggr ) =
\Psi \Biggl ( {1 \over 2} \Biggr ) -
\Psi \Biggl ( {1 \over 2} + {{1-\bar V_{l_0,l_0}} \over
{4\pi \tau T_c^{\rm dis}}} \Biggr ) \,.
\end{equation}
For $T_c-T_c^{\rm dis}\ll T_c$, the
preceding equation simplifies in
\begin{equation}
{{T_c-T_c^{\rm dis}} \over {T_c}} \simeq {\pi \over {8\tau T_c}}
(1-\bar V_{l_0,l_0}) \,.
\label{Tcdis}
\end{equation}
In the preceding equation, $T_c$ is the critical temperature calculated in
the QLA without impurity scattering. Since $T_c$ is several orders of
magnitude below the exact mean-field value, $T_c^{\rm dis}$ will also be
much smaller than the exact value. However, a reasonable estimation of
$T_c^{\rm dis}$ can be obtained using for $T_c$ the exact value instead of
the QLA value. From Eq.\ (\ref{Tcdis}), one can see that
$\bar V_{l_0,l_0}$, which comes from vertex corrections in the pair
propagator, tends to reduce the difference $T_c-T_c^{\rm dis}$. According to
(\ref{Tcdis}), impurity scattering does not affect the critical temperature
in the reentrant phase since $\bar V_{l_0,l_0}\rightarrow 1$ when $T_c
\rightarrow T_{c0}$. This is a direct consequence of Anderson's theorem
which
states that the critical temperature is independent of a (weak) disorder for
a system with time-reversal symmetry. \cite{Anderson59} Obviously,
(\ref{Tcdis}) restricts the observation of high-field
superconductivity to clean superconductors with a critical temperature not
too small. As pointed out in Ref. \cite{Dupuis94a}, this latter condition
advantages materials with a large anisotropy. The consequences
of (\ref{Tcdis}) were discussed in detail in Ref. {\cite{Dupuis94a}
in the case of the Bechgaard salts.

\subsection{Effect of PPB}
\label{SubSecEPPB}

We evaluate the effect of PPB on the critical temperature calculated in
Sec.\ \ref{SubSecWDWPPB} (but ignoring the effect of disorder). The equation
(\ref{EQVertex1}) for the two-particle vertex function $\Gamma
_{q_x}^{\alpha \alpha '}(l_1,l_2;l'_1l'_2)$ involves the quantity $\chi
^{\alpha ''}(q_x+(l''_1+l''_2)G)$ which is given by (\ref{PP0})
with the replacement $\alpha vq_x \rightarrow \alpha vq_x+2h$.
Therefore, logarithmic singularities arise through $\chi  _{q_x}^{\alpha ''}
(q_x+(l''_1+l''_2)G)$ each time we have $q_x=-2L''G+q_0$
where $L''$ is the center of gravity of the pair in an
intermediate state and $q_0=\pm 2h/v$. In the QLA, we retain only the
intermediate states corresponding to these logarithmic singularities. If
$h/\omega _c$ is not (and not too close to) an integer, the center
of gravity $L$ of the pair becomes a constant of motion and is related to
the total momentum by $q_x(L)=-2LG+q_0$. The equation which determines
the two-particle vertex function then reduces to
\begin{equation}
\Gamma ^{\alpha \alpha '}_{q_x(L),L}(l,l')=-\lambda
V_{l,l'}^{\alpha \alpha '} +\lambda \sum _{\alpha '',l''}
V_{l,l''}^{\alpha \alpha ''} \chi ^{\alpha ''}(q_0)
\Gamma ^{\alpha ''\alpha '}_{q_x(L),L}(l'',l')  \,.
\end{equation}
Following the analysis developed in Sec.\ \ref{SubSecWDWPPB}, we find that
a pole appears in the two-particle vertex function when
\begin{equation}
1-\lambda \bar V_{l,l} \chi (q_0)=0  \,.
\end{equation}
Using
\begin{equation}
\chi (q_0)\simeq {{N(0)} \over 2} \ln \Biggl ( {{\gamma \Omega ^2}
\over {\pi Th}} \Biggr )
\end{equation}
for $h\gg T$, we obtain the critical temperature
\begin{eqnarray}
T_c^{\rm P} &\simeq & {{\gamma \Omega ^2} \over {\pi h}}
\exp \Biggl ( {{-2} \over {\lambda N(0)\bar V_{l_0,l_0}}} \Biggr ) \nonumber
\\ &\simeq & {{\pi T_c^2} \over {4\gamma h}} \,,
\label{TcPPB}
\end{eqnarray}
where  $\bar V_{l_0,l_0}={\rm max}_l V_{l,l}$. In the reentrant phase
($\omega _c \gg t_z$), $T_c \rightarrow T_{c0}$ so that $T_c^{\rm P}
\rightarrow \pi T_{c0}^2/4\gamma h$, a result which was obtained in
Ref. \cite{Dupuis94a}. The superconducting phase is always stable at $T=0$,
whatever the value of the magnetic field. This divergence of the orbital
critical field $H_{c2}(T)$ at low temperature is a characteristic of the
LOFF state in a quasi-1D superconductor. \cite{Dupuis94a,Dupuis94c}

If $h/\omega _c$ is equal (or close) to an integer, the preceding analysis
does not hold any more. In this case, the center of gravity of the Cooper
pair can take the two values $-q_x/2G\pm h/\omega _c$. This enlarges the
number of accessible intermediate states and should lead to an increase of
the critical temperature with respect to the result (\ref{TcPPB}). It has
been proposed by Lebed' \cite{LebedU} that this situation could be reached
by tilting the field in the $(x,y)$ plane. This lets the Zeeman energy $h$
unchanged but modifies the orbital frequency which becomes $\omega _c\cos
\theta $ where $\theta $ is the angle between the field and the $y$ axis.
For certain values of $\theta $, $h/(\omega _c\cos \theta )$ is an
integer. The effect of the component $H_x=H\sin \theta $ is expected to be
small (if $\theta $ is not too large), because the critical field
$H_{c2}^x(0)$ parallel to the $x$ axis is very large compared to the
critical fields in the other directions.

\subsection{Effect of PPB and disorder}
\label{SubSecEPPBD}

We evaluate the effects of both PPB and disorder on the
critical temperature calculated in Sec.\ \ref{SubSecWDWPPB}. Following the
analysis developed in the two preceding sections, we find that the critical
temperature $T_c^{\rm P,dis}$ is determined by the equation
\begin{equation}
N(0) \ln \Biggl ( {T \over {T_c^{\rm P}}} \Biggr ) =
T \sum _\omega  {{\chi _{\tilde \omega }(q_0)} \over
{1-u_0\bar V_{l_0,l_0} \chi _{\tilde \omega }(q_0) }}
-T \sum _\omega  \chi _\omega (q_0) \,,
\label{Eq40}
\end{equation}
where $T_c^{\rm P}$ is the critical temperature calculated in the preceding
section. For $T_c^{\rm P}-T_c^{\rm P,dis} \ll T_c^{\rm P}$, we can expand
(\ref{Eq40}) to first order in $u_0$ (or $1/\tau $), which leads to
\begin{equation}
N(0) \ln \Biggl ( {T \over {T_c^{\rm P}}} \Biggr ) \simeq
T \sum _\omega \lbrack \chi _{\tilde \omega }(q_0)
-\chi _\omega (q_0) \rbrack +u_0 \bar V_{l_0,l_0}
T \sum _\omega  \chi _\omega ^2(q_0)  \,,
\end{equation}
where
\begin{equation}
\chi _\omega (q_0) = {i \over {vc}} {\rm sgn}(\omega )
\Biggl \lbrack {1 \over {2i\omega }} + {1 \over {2i\omega -4h}}
\Biggr \rbrack \,,
\end{equation}
\begin{equation}
\chi _{\tilde \omega }(q_0)-\chi _\omega (q_0)\simeq
- {{\pi N(0)} \over \tau } \Biggl \lbrack {1 \over {4\omega ^2}}
-{1 \over {(2i\omega -4h)^2}} \Biggr \rbrack \,.
\end{equation}
Performing the sum over the Matsubara frequencies, we obtain
\begin{equation}
\ln \Biggl ( {T \over {T_c^{\rm P}}} \Biggr ) \simeq
-{\pi \over {16\tau T}} \Biggl ( 1-{{\bar V_{l_0,l_0}} \over 2} \Biggr )
\Biggl \lbrack 1+ {2 \over {\pi ^2}} {\rm Re}\, \Psi ' \Biggl ( {1 \over 2}
+i{h \over {\pi T}} \Biggr ) \Biggr \rbrack
+{{\bar V_{l_0,l_0}} \over {8\tau h}} {\rm Im}\, \Psi \Biggl ( {1 \over 2}
+i{h \over {\pi T}} \Biggr ) \,,
\end{equation}
where $\Psi '$ is the first derivative of the digamma function. For $h \gg
\pi T$, we can use $\Psi (z)\simeq \ln z-1/2z$ for $\vert z \vert \gg 1$ to
get the following expression to lowest order in $1/\tau $:
\begin{equation}
{{T_c^{\rm P,dis}-T_c^{\rm P}}  \over {T_c^{\rm P}}}
\simeq -{\pi \over {16\tau T_c^{\rm P}}}
\Biggl ( 1-{{\bar V_{l_0,l_0}} \over 2} \Biggr )
+{\pi \over {16\tau T_c^{\rm P}}}  \Biggl ( 1+{{\bar V_{l_0,l_0}} \over 2}
\Biggr ) {{{T_c^{\rm P}}^2} \over {h^2}}  \,.
\label{TcPPBdis}
\end{equation}
As in Sec.\ \ref{SubSecEOD}, a reasonable estimation of the effect of
disorder can be obtained from the preceding equation using for $T_c^{\rm P}$
the exact mean-field value.
The effect of vertex corrections is weaker than in the absence of PPB
(Eq.\ (\ref{Tcdis})). Even in the reentrant phase where $\bar V_{l_0,l_0}
\rightarrow 1$, the effect of impurities remains important since
time-reversal symmetry is broken by the Zeeman splitting and Anderson's
theorem does not hold. The sensitivity of the LOFF state to elastic
impurity scattering has been known for a long time. \cite{Aslamazov69}
The consequences of (\ref{TcPPBdis}) were discussed in detail in
Ref. \cite{Dupuis94a} in the case of the Bechgaard salts.

\section{Ordered phase in the quantum limit approximation}
\label{SecOPQLA}

In this section, we study the ordered phase ($T<T_c$) in the QLA where only
the pairing channels leading to Cooper logarithmic singularities are
retained. We only
consider the orbital effects of the magnetic field, i.e. we put the $g$
factor equal to zero.
Using the results obtained in Sec.\ \ref{SubSecWDWPPB}, we first
construct a variational order parameter with two unknown parameters,
the amplitude and the periodicity in the $z$ direction.
We solve the Gor'kov equations in the QLA and obtain
the normal and anomalous Green's functions. We then deduce
the GL expansion of the free energy, the specific
heat jump at the transition and the quasi-particle excitation spectrum.

Following the original approach proposed by Abrikosov,
\cite{Abrikosov57,Parks69}
we construct the order parameter for $T<T_c$ as a linear combination of the
functions $\Delta _{q_x(L),L,l_0}({\bf r})$ (Eq.\ (\ref{po1})) describing the
superconducting condensation in the channel $q_x(L),L,l_0$. Since $T_c$ does
not depend on $L$, the most general linear combination can be written as
\begin{equation}
\Delta ({\bf r})= \sum _L \gamma (L) \Delta _{q_x(L),L,l_0}({\bf r}) \,.
\end{equation}
As shown in Sec.\ \ref{SubSecWDWPPB}, $2L$ must have the parity of $l_0$.
Since $\Delta _{q_x(L),L,l_0}({\bf r})$ is localized in the $z$ direction
with an  extension of $c\tilde t$, a natural choice for the coefficients
$\gamma (L)$ is to take $\gamma (L) \neq 0$ if $L=-l_0/2+pN'$ ($p$ integer)
where the unknown integer $N'$ is expected to be of order $\tilde t$. In
order to
correctly describe the triangular Josephson vortex lattice in the last phase
(which has periodicity $2c$ in the $z$ direction), \cite{Dupuis94a} we
choose $\gamma (L)\equiv \gamma _p=1\,(i)$ for $p$ even (odd). This leads
to (noting $N=2N'$)
\begin{equation}
\Delta _{l_0,N}({\bf r})=\Delta \sum _{l,p} U_{2l+l_0,l_0} \gamma _p
e^{i(l_0-pN)Gx} J_{p{N \over 2}+l-m}(\tilde t)
J_{p{N \over 2}-l-l_0-m}(-\tilde t)  \,,
\label{po2}
\end{equation}
where the amplitude $\Delta $ is chosen real.
Eq.\ (\ref{po2}) defines a
variational order parameter where the two unknown parameters $\Delta $ and
$N$ have to be determined by minimizing the free energy. It can be seen
that $\vert \Delta ({\bf r})\vert $ has periodicity $a_x=2\pi /NG$ and
$a_z=Nc$ so that the unit cell contains two flux quanta $\phi _0$ for a
particle of charge $2e$: $Ha_xa_z=2\phi _0$
(when a triangular lattice is described with a square unit cell, the unit
cell contains two flux quanta). In Ref. \cite{Dupuis94a}, the order parameter
was constructed by imposing that it describe both the triangular Abrikosov
vortex lattice in weak field ($\omega _c\ll T$) and the triangular Josephson
vortex lattice in very strong field ($\omega _c \gg t_z$). Both approaches
lead to the same order parameter when only the Cooper singularities are
retained.

The order parameter is shown in Figs.\ \ref{FigPON2}-\ref{FigPON10},
for the phases $N=2,4,6,8$ and 10.
The exact (at the mean-field level) order parameter corresponding
to Figs.\ \ref{FigPON2}, \ref{FigPON6}, and \ref{FigPON8}, is shown in figures
10, 9 and 8 of Ref. \cite{Dupuis94a}. The order parameter calculated in the
QLA is
a good approximation of the exact order parameter for the last phases $N=2$,
4 and 6. For larger values of $N$, there appears significant differences
between the approximate and exact results. For example, in the phase
$N=8$ at $H=1.7$ T, the order parameter obtained in the QLA shows zeros
which are not present in the exact calculation.

\subsection{Gor'kov equations }
\label{SubSecGEES}

The starting point of our analysis is the mean-field  (or generalized
Hartree-Fock) Hamiltonian ${\cal
H}_0+{\cal H}_{\rm int}^{\rm MF}$ with
\begin{equation}
{\cal H}_{\rm int}^{\rm MF}= {1 \over 2} \int d^2{\bf r} \,
\sum _{\alpha ,\sigma } \Bigl \lbrack \Delta _\sigma ({\bf r})^*
\psi _{\overline \sigma }^{\overline \alpha }({\bf r})
\psi _\sigma ^\alpha ({\bf r}) +{\rm H.c.}
\Bigr \rbrack \,,
\end{equation}
where
\begin{equation}
\Delta _\sigma ^*({\bf r}) = \lambda \sum _\alpha
\langle \psi _{\overline \sigma }^{\overline \alpha  \dagger }({\bf r})
\psi _\sigma ^{\alpha \dagger }({\bf r}) \rangle \,.
\end{equation}
$\Delta _\uparrow ({\bf r})= -\Delta _\downarrow ({\bf r})$ is the
variational order parameter defined by (\ref{po2}).
In order to derive the thermodynamics and the excitation spectrum in the
superconducting phases, it is necessary to determine the normal and
anomalous Green's functions
\begin{eqnarray}
G_\sigma ^\alpha ({\bf r},{\bf r}',\tau ) &=&
-\langle T_\tau \psi _\sigma ^\alpha ({\bf r},\tau ) \psi _\sigma ^{\alpha
\dagger }({\bf r},0) \rangle \,, \nonumber \\
F_\sigma ^{\alpha \dagger }({\bf r},{\bf r}',\tau ) &=&
-\langle T_\tau \psi _\sigma ^{\alpha \dagger }({\bf r},\tau )
\psi _{\overline \sigma }^{\overline \alpha \dagger }({\bf r},0) \rangle \,,
\end{eqnarray}
whose Fourier transforms with respect to the imaginary time $\tau $
are solutions of the Gor'kov equations
\begin{eqnarray}
& & (i\omega -{\cal H}_{0,\sigma }^\alpha ) G_\sigma ^\alpha ({\bf r},{\bf
r}',\omega ) -\Delta _\sigma ({\bf r})
F_{\overline \sigma }^{\overline \alpha \dagger }({\bf r},{\bf r}',\omega )
= \delta ({\bf r}-{\bf r}') \,, \nonumber \\
& & (-i\omega -{\cal H}_{0,\overline \sigma }^{\overline \alpha \dagger })
F_{\overline \sigma }^{\overline \alpha \dagger }({\bf r},{\bf r}',\omega )
+\Delta _\sigma ^*({\bf r})
G_\sigma ^\alpha ({\bf r},{\bf r}',\omega ) = 0 \,,
\label{GE1}
\end{eqnarray}
with the self-consistency equation
\begin{equation}
\Delta _\sigma ^*({\bf r}) =
\lambda T \sum _{\alpha ,\omega }
F_\sigma ^{\alpha \dagger }({\bf r},{\bf r},\omega ) \,.
\label{SCE1}
\end{equation}
In the QLA, the
supercurrents vanish as will be shown in Sec.\ \ref{SecBQLA}. The
magnetization
$M$ (which is parallel to the magnetic field by symmetry arguments) and the
flux density $B=H+4\pi M$ are uniform. In the Gor'kov equations, we should
therefore in principle replace the external magnetic field $H$ by the
uniform flux density $B$. However, when minimizing the free energy with
respect to $B$, it turns out that the approximation $B=H$ is very accurate
(see Sec.\ \ref{SubSecGLET}). Moreover, the magnetization vanishing in the
middle
of each phase (see below), this approximation becomes exact at this point.
In this section, we consider that the flux density is equal to the external
magnetic field.

To exploit the translational symmetry of the order parameter, we
introduce the magnetic Bloch states $\phi _{{\bf q},l}^\alpha $ obtained by
diagonalizing simultaneously the three operators ${\cal H}_0$, $e^{a_x
\partial _x}$ and $e^{a_z \partial _z +iNGx}$ which define the magnetic
translation group:
\begin{equation}
\phi _{{\bf q},l}^\alpha = \sqrt{{N \over {N_z}}}
\sum _p e^{-ipq_za_z} \phi _{q_x+pNG,l-pN}^\alpha \,.
\label{BLOCH}
\end{equation}
These states are eigenstates of the Hamiltonian ${\cal H}_{0,\sigma }^\alpha
$  with the eigenenergies
$\epsilon _{{\bf q},l,\sigma }^\alpha =\epsilon _{q_x,l,\sigma }^\alpha $.
$N_z$ is the number of chains. ${\bf q}$
is restricted to the magnetic Brillouin zone (MBZ) which is chosen to be
equal to
\begin{equation}
\Biggl \lbrack -k_F-l_0G-{\pi \over {a_x}},-k_F-l_0G+{\pi \over {a_x}} \Biggr
\lbrack \times  \Biggl \lbrack -{\pi \over {a_z}},{\pi \over {a_z}} \Biggr
\lbrack \,
\label{MBZ1}
\end{equation}
for the left sheet of the Fermi surface, and to
\begin{equation}
\Biggl \rbrack k_F-{\pi \over {a_x}},k_F+{\pi \over {a_x}} \Biggr \rbrack
\times \Biggl \rbrack -{\pi \over {a_z}},{\pi \over {a_z}} \Biggr \rbrack
\label{MBZ2}
\end{equation}
for the right sheet of the Fermi surface. This choice is made for later
convenience. There are $N$ branches crossing the Fermi level on each sheet
of the Fermi surface. The excitation spectrum of the normal phase in the
representation of the magnetic Bloch states is shown
in Fig.\ \ref{FigBlochSpec}. We introduce Green's functions in the
representation of these magnetic Bloch states writing
\begin{eqnarray}
G_\sigma ^\alpha ({\bf r},{\bf r}',\omega ) &=& \sum _{1,2}
\phi _1^\alpha ({\bf r}) \phi _2^\alpha ({\bf r}')^*
G_\sigma ^\alpha (1,2,\omega ) \,, \nonumber \\
F_\sigma ^{\alpha \dagger }({\bf r},{\bf r}',\omega ) &=& \sum _{1,2}
\phi _1^\alpha ({\bf r})^* \phi _2^{\overline \alpha }({\bf r}')^*
F_\sigma ^{\alpha \dagger }(1,2,\omega ) \,,
\end{eqnarray}
where we use the notation $i\equiv ({\bf q}_i,l_i)$. In this representation,
the Gor'kov equations become
\begin{eqnarray}
& & (i\omega -\epsilon ^\alpha _{1\sigma } )
G^\alpha _\sigma (1,2,\omega ) - \sum _3 \Delta _\sigma ^\alpha (1,3)
F^{\overline \alpha \dagger}_{\overline \sigma }(3,2,\omega ) =\delta _{1,2}
\,, \nonumber \\   & &
(-i\omega -\epsilon ^{\overline \alpha }_{1\overline \sigma } )
F^{\overline \alpha \dagger}_{\overline \sigma }(1,2,\omega )
+\sum _3 \Delta _\sigma ^{\overline \alpha }(1,3)^*
G^\alpha _\sigma (3,2,\omega )  =0 \,,
\label{GE2}
\end{eqnarray}
where the pairing amplitude $\Delta ^\alpha _\sigma (1,2)$ is defined by
\begin{equation}
\Delta ^\alpha _\sigma (1,2)= \int d^2{\bf r}\,
\phi _1^\alpha ({\bf r})^* \phi _2^{\overline \alpha }({\bf r})^*
\Delta _\sigma ({\bf r}) \,.
\label{Ampli}
\end{equation}
Multiplying (\ref{SCE1}) by $\Delta _\sigma ({\bf r})$ and
summing over ${\bf r}$, we can rewrite the self-consistency equation as
\begin{equation}
\int d^2{\bf r} \, \vert \Delta _\sigma ({\bf r}) \vert ^2 =
\lambda T \sum _{\omega ,\alpha } \sum _{1,2}
\Delta _\sigma ^\alpha (1,2) F_\sigma ^{\alpha \dagger } (1,2,
\omega )  \,.
\label{SCE2}
\end{equation}

A careful examination shows that the pairing between
$\phi ^\alpha _{{\bf q},l}$ and $\phi ^{\overline \alpha }_{l_0{\bf G}-{\bf
q},-l_0-l}$ (with ${\bf G}=(G,0)$) is the only pairing compatible with the
spatial
dependence of the order parameter (\ref{po2}). Thus, in the QLA, we have
\begin{equation}
\Delta _\sigma ^\alpha ({\bf q}_1,l_1;{\bf q}_2,l_2)
= \delta _{{\bf q}_1+{\bf q}_2,l_0{\bf G}} \delta _{l_1+l_2,-l_0}
\Delta _\sigma ^\alpha ({\bf q}_1,l_1) \,.
\label{CondQLA}
\end{equation}
The preceding result can also be obtained noting that in the QLA the
electron-electron interaction conserves the center of gravity of the
particles with respect to the $z$ direction (Sec.\ \ref{SecTL}).
Eq.\ (\ref{CondQLA}) follows from this property of the interaction.
The definition of the MBZ (\ref{MBZ1},\ref{MBZ2}) ensures that if
${\bf q}$ belongs to the MBZ, then $l_0{\bf G}-{\bf q}$ also belongs to the
MBZ. The pairing amplitude $\Delta _\sigma ^\alpha ({\bf q},l)$ can be
calculated using (\ref{Estates},\ref{po2},\ref{BLOCH}):
\begin{eqnarray}
\Delta _\sigma ^\alpha ({\bf q},l)&\equiv &
\Delta _\sigma ^\alpha (q_z,l) \nonumber \\ &=&
\Delta _\sigma \alpha ^{l_0} \bar V_{l_0,l_0}
\sum _p \gamma _p e^{-ipq_za_z} U_{2l+l_0+pN,l_0} \,,
\label{AmpliQLA}
\end{eqnarray}
and $\Delta _\uparrow =-\Delta _\downarrow =\Delta $. The equality
(\ref{CondQLA}) and the Gor'kov equations (\ref{GE2}) imply that the
Green's functions in the representation of the magnetic Bloch states are
diagonal:
\begin{eqnarray}
G^\alpha _\sigma ({\bf q}_1,l_1;{\bf q}_2,l_2,\omega )&=&\delta _{{\bf
q}_1,{\bf q}_2} \delta _{l_1,l_2} G^\alpha _\sigma ({\bf
q}_1,l_1,\omega ) \,, \nonumber \\
F^{\alpha \dagger }_\sigma ({\bf q}_1,l_1;{\bf q}_2,l_2,\omega )&=&\delta
_{{\bf q}_1+{\bf q}_2,l_0{\bf G}} \delta _{l_1+l_2,-l_0}
F^{\alpha \dagger }_\sigma ({\bf q}_1,l_1,\omega ) \,,
\end{eqnarray}
where $G_\sigma ^\alpha ({\bf q},l,\omega )$ and $F_\sigma ^{\alpha \dagger
}({\bf q},l,\omega )$ are given by:
\begin{eqnarray}
G_\sigma ^\alpha ({\bf q},l,\omega )&=&
{{-i\omega -\epsilon  _{{\bf q},l,\sigma }^\alpha } \over
{\omega ^2+{\epsilon  _{{\bf q},l,\sigma }^\alpha }^2 + \vert \Delta _\sigma
^\alpha ({\bf q},l) \vert ^2 }} \,,  \nonumber \\
F_\sigma ^{\alpha \dagger }({\bf q},l,\omega )&=&
{{\Delta _\sigma ^{\alpha }({\bf q},l)^* }  \over
{\omega ^2+{\epsilon  _{{\bf q},l,\sigma }^\alpha }^2 + \vert \Delta _\sigma
^\alpha ({\bf q},l) \vert ^2 }}  \,.
\label{GEQLA}
\end{eqnarray}
In the next sections, we will use the preceding expressions of the Green's
functions to obtain the GL expansion of the free energy and the excitation
spectrum.

\subsection{Ginzburg-Landau expansion and thermodynamics }
\label{SubSecGLET}

We first note that the sign of the magnetization in the ordered phase can be
obtained from general thermodynamics arguments. \cite{Montambaux89}
The magnetization is obtained from $M=-\partial F_e/\partial B$, where
$F_e(T,B)$ is the electronic contribution to the difference of the free
energies of the normal and
superconducting phases. Since $T_c$ is determined
by $F_e(T_c,B)=0$, the magnetization $M$ close to the transition line has the
sign of $dT_c/dH$.
Each phase will therefore first be paramagnetic and then diamagnetic for
increasing field (which implies that there is a value of the field in each
phase for which the magnetization vanishes), except the reentrant phase
which is always paramagnetic. \cite{Melo}

Since the flux density $B$ is uniform in the QLA,
the free energy density is equal to $F(T,B)=F_e(T,B)+B^2/8\pi $
where the electronic contribution can be written \cite{Fetter71}
\begin{eqnarray}
F_e(T,B)&=&\int _0^\Delta d\Delta ' {{dg} \over {d\Delta '}} {\Delta '}^2
\int {{d^2{\bf r}} \over S} \Biggl \vert
{{ \Delta _\sigma ({\bf r})} \over \Delta } \Biggr \vert ^2
\nonumber \\
&=& 2\int _0^\Delta \Delta 'd\Delta '(\lambda ^{-1}-g(\Delta '))
\int {{d^2{\bf r}} \over S} \Biggl \vert
{{ \Delta _\sigma ({\bf r})} \over \Delta } \Biggr \vert ^2 \,,
\label{freeFE}
\end{eqnarray}
where the function $g(\Delta )=1/\lambda $ is defined by (\ref{SCE2}).
In (\ref{freeFE}), all quantities are calculated in presence of a constant
flux density $B$. In order to get the free energy per volume unit (and not
per surface unit), the 2D density of states $N(0)=1/\pi vc$ should be
replaced by its 3D expression $1/\pi vbc$.
Expanding $F_\sigma ^{\alpha \dagger }$ in (\ref{SCE2})
in power of $\Delta $ leads to
\begin{equation}
F_e(T,B) = \alpha \Delta ^2 +{\beta \over 2} \Delta ^4 \,,
\label{freeFE1}
\end{equation}
with
\begin{eqnarray}
\alpha  &=& {1 \over \lambda } \int {{d^2{\bf r}} \over S} \Biggl
\vert {{ \Delta _\sigma ({\bf r})} \over \Delta } \Biggr \vert ^2
-{T \over S} \sum _{\alpha ,\omega ,{\bf q},l}
{{\vert \Delta _\sigma ^\alpha ({\bf q},l) / \Delta  \vert ^2  }
\over {\omega ^2+{\epsilon  _{{\bf q},l,\sigma }^\alpha }^2 }}
\,, \nonumber \\
\beta &=& {T \over S} \sum _{\alpha ,\omega ,{\bf q},l}
{{\vert \Delta _\sigma ^\alpha ({\bf q},l) / \Delta  \vert ^4  }
\over {(\omega ^2+{\epsilon  _{{\bf q},l,\sigma }^\alpha }^2)^2 }}  \,.
\end{eqnarray}
In the two preceding equations (and in the following), the sum over ${\bf q}$
is restricted to the MBZ. Using (\ref{po2},\ref{AmpliQLA}), and the
property
\begin{equation}
\Delta _\sigma ^\alpha (q_z,l+pN)=
e^{2ipq_za_z} \Delta _\sigma ^\alpha (q_z,l) \,,
\end{equation}
we obtain
\begin{eqnarray}
\alpha  &=& {{2\bar V_{l_0,l_0}} \over N} (\lambda ^{-1}-\bar
V_{l_0,l_0} \chi (0)) \nonumber \\
&\simeq & 2N(0){{\bar V_{l_0,l_0}^2} \over N}
{{T-T_c} \over {T_c}} \,, \nonumber \\
\beta  &=& \beta _{\rm BCS} {{\bar V_{l_0,l_0}^4} \over N} \sum
_{l=1}^N \sum _{p_1,p_2,p_3} \gamma _{p_1} \gamma _{p_2}
\gamma _{p_3}^* \gamma _{p_1+p_2-p_3}^*
U_{2l+l_0+p_1N,l_0} \nonumber \\ & & \times U_{2l+l_0+p_2N,l_0}
U_{2l+l_0+p_3N,l_0} U_{2l+l_0+(p_1+p_2-p_3)N,l_0} \,,
\label{GLE2}
\end{eqnarray}
where $\beta $ is calculated at $T_c$ and
$\beta _{\rm BCS}=7\zeta (3)N(0)/8\pi ^2T_c^2$. The equilibrium state is
determined by the minimum of the Gibbs free energy
\begin{eqnarray}
G(T,H)&=&F(T,B)-{{BH} \over {4\pi }} \nonumber \\
&=& F_e(T,B) +{{(B-H)^2} \over {8\pi }} -{{H^2} \over {8\pi }} \,.
\end{eqnarray}
Minimizing $G(T,H)$ (or equivalently $F_e(T,B)$) with respect to $\Delta $,
we obtain
\begin{equation}
F_e(T,B)=-{{\alpha ^2} \over {2\beta }} \,.
\label{freeFE2}
\end{equation}
Since $\alpha $ and $\beta $ are known, we can minimize numerically $G(T,H)$
with respect to $B$ and then look for the integer $N$ which minimizes the
Gibbs free energy. It turns out that the minimum is reached
for $B=H$ with a very high accuracy. As pointed out above, this approximation
becomes exact in the middle of each phase where the magnetization vanishes.
In the following, we therefore consider
that the flux density is equal to the external magnetic field. Thus, we can
consider the free energy $F_e(T,H)$ instead of the Gibbs free energy
$G(T,H)$. In the reentrant phase, we find that
the minimum of $F_e(T,H)$ is obtained for $N=2$.
When the field is decreased from its value in the reentrant phase, the
system undergoes a first order transition and the minimum of $F_e$ is then
obtained for $N=4$. This is in agreement with Refs. \cite{Dupuis93,Dupuis94a}
where it is argued that the first order phase transitions are due to
commensurability effects between the crystalline lattice spacing and the
periodicity of the order parameter. Unlike what is expected, the best value
of $N$ switches to 6 before reaching the next first order transition. This
indicates that the QLA gives a bad estimate of the free energy in the phases
$N\geq 4$. A correct calculation of the free energy in these phases requires
the inclusion of all the pairing channels and maybe also of the screening of
the magnetic field. This point will be further discussed in Sec.\
\ref{SubSecBQLAGLE}.

The calculation of the free energy can easily be extended to the case of a
more general order parameter defined by (\ref{po2}) but with $\gamma _p=1$
($\gamma $) for $p$ even (odd) where $\gamma $ is an unknown parameter.
For a fixed periodicity $N$, we
have found that the free energy is stationary with respect to $\gamma $ for
$\gamma =0$, $1$ or $i$ which corresponds to square lattices of
periodicities $a_z=Nc$ and $a_z=Nc/2$, and to a triangular lattice of
periodicity $a_z=Nc$. Numerically, we have verified that in the reentrant
phase, the triangular lattice with periodicity $N=2$ has a lower free energy
than the square lattice with periodicity $N=1$ or $N=2$.
This result (stationarity of the free energy for the square and triangular
lattices, the latter corresponding to the minimum of the free energy)
extends to the quantum regime $\omega _c\gg T$ a property which is
well-known in the GL regime. \cite{Parks69,Kleiner64}

The specific heat jump at the transition is obtained from $\Delta
C=-T\partial ^2F_e/\partial T^2$. From (\ref{GLE2},\ref{freeFE2}),
we obtain
\begin{eqnarray}
r&=&{{\Delta C/C_N} \over {(\Delta C/C_N)_{\rm BCS}}} \nonumber \\
&=& 4 \Biggl \lbrack
N \sum _{l=1}^N \sum _{p_1,p_2,p_3} \gamma _{p_1} \gamma _{p_2}
\gamma _{p_3}^* \gamma _{p_1+p_2-p_3}^*
U_{2l+l_0+p_1N,l_0} \nonumber \\ & & \times U_{2l+l_0+p_2N,l_0}
U_{2l+l_0+p_3N,l_0} U_{2l+l_0+(p_1+p_2-p_3)N,l_0} \Biggr \rbrack ^{-1}\,,
\end{eqnarray}
where $C_N$ is the specific heat of the normal state and $(\Delta
C/C_N)_{\rm BCS}=12/7\zeta (3)$ is the zero-field value. The ratio $r$
is always smaller than 1 (but close to 1 in the reentrant phase) and
discontinuous at each first order phase transition (Fig.\ \ref{FigSH}). As
in the case of the FISDW phases, \cite{Poilblanc88} these discontinuities
can be related to the slopes $\Delta T/\Delta H$ of the first order
transition lines. Consider two consecutive superconducting phases denoted by
1 and 2. For each phase $i=1,2$, the free energy can be written as
($T\leq T_c^{(i)}$)
\begin{equation}
F_i(T,H)=-a_i(T_c^{(i)},H)(T-T_c^{(i)})^2 \,,
\end{equation}
and the specific heat jump at the transition is $\Delta
C_i=2a_i(T_c^{(i)},H)T_c^{(i)}$. $T_c^{(i)}$ denotes the transition
temperature between the normal phase and the superconducting phase $i$.
The first order transition line
is obtained by writing the equality of the free energy between the two phases:
\begin{equation}
-a_1(T_c^{(1)},H)(T-T_c^{(1)})^2 = -a_2(T_c^{(2)},H)(T-T_c^{(2)})^2 \,.
\end{equation}
Introducing the quantities $\Delta T=T-T^*$ and $\Delta H=H-H^*$ where the
point
$(H^*,T^*)$ is defined by the intersection of the two transition lines, i.e.
$T^*=T_c^{(1)}(H^*)=T_c^{(2)}(H^*)$, the preceding equality is rewritten as
\begin{equation}
\sqrt{a_1(T_c^{(1)},H)} \Biggl ( {{\Delta T} \over {\Delta H}} -
{{\Delta T_c^{(1)}} \over {\Delta H}} \Biggl ) =
\sqrt{a_2(T_c^{(2)},H)} \Biggl ( {{\Delta T} \over {\Delta H}} -
{{\Delta T_c^{(2)}} \over {\Delta H}} \Biggl )  \,.
\end{equation}
Taking the limit $T\rightarrow T^*$ and $H\rightarrow H^*$, we obtain
\begin{equation}
{{\Delta T} \over {\Delta H}} =
{ { \sqrt{\Delta C_1}{{dT_c^{(1)}} \over {dH}} -\sqrt{\Delta C_2}{{dT_c^{(2)}}
\over {dH}} } \over {\sqrt{\Delta C_1}-\sqrt{\Delta C_2}}} \,,
\label{DTDH}
\end{equation}
where $\Delta C_1\equiv \Delta C_1(H^*)$ and  $\Delta C_2\equiv \Delta
C_2(H^*)$. Since $dT_c^{(1)}/dH<0$ and $dT_c^{(2)}/dH>0$ (Fig.\
\ref{FigTCQLA}), it is clear that
the slope $\Delta T/\Delta H$ of the first
order transition line has the sign
of $\Delta C_2-\Delta C_1$: the slope is positive for the transition between
the phases $N=4$ and $N=2$ and negative for the other transitions.

The calculation of the specific heat jump at the transition clearly shows an
important difference between the study of the superconducting phases in the
QLA and the study of the FISDW phases in the SGA.
\cite{Virosztek86,Montambaux88,Maki86} In the SGA, the electron-hole pairing
amplitudes are given by a single number (corresponding to the gap which
opens at the Fermi level) so that the thermodynamics
of the FISDW reduces to that of the (zero-field) BCS model. For example, one
obtains $r=1$ for any value of the magnetic field. In order to
get corrections to the BCS description, one has to go beyond the SGA. In the
QLA used here for the study of the superconducting phases, the
(electron-electron) pairing amplitudes $\Delta ^\alpha _\sigma (q_z,l)$ do
not reduce to a single number so that the thermodynamics is different from
that of the (zero-field) BCS model.

\subsection{Excitation spectrum}
\label{SubSecESQLA}

{}From the Green's functions (\ref{GEQLA}), we deduce the quasi-particle
spectrum in the QLA:
\begin{eqnarray}
E_{{\bf q},l,\sigma }^\alpha &=& \pm \sqrt{ {\epsilon _{{\bf
q},l,\sigma }^\alpha }^2
+\vert \Delta _\sigma ^\alpha ({\bf q},l)\vert ^2 }
\nonumber \\ &=& \pm \sqrt{ {\epsilon _{q_x,l,\sigma }^\alpha }^2
+\vert \Delta _\sigma ^\alpha (q_z,l)\vert ^2 } \,.
\label{Exci}
\end{eqnarray}
For each branch $l$ which crosses the Fermi level, a gap $2 \vert \Delta
_\sigma ^\alpha (q_z,l) \vert $ opens at the Fermi level for
$q_x=\alpha k_F-lG$ where $\epsilon ^\alpha _{q_x,l,\sigma }=0$. The pairing
amplitudes $\Delta ^\alpha _\sigma (q_z,l)$ are shown in Fig.\
\ref{FigPAQLA} for the last five phases $N=2$, $N=4$, $N=6$, $N=8$ and
$N=10$. Note that $\vert \Delta _\sigma ^\alpha (-q_z,l\pm N/2)\vert =
\vert \Delta _\sigma ^\alpha (q_z,l)\vert $.
In the very high field limit $\tilde t\ll 1$, the dispersion of
$\Delta ^\alpha _\sigma (q_z,l)$ with respect to $q_z$ is very weak (of
the order of $\tilde t^2$). The corresponding excitation spectrum for the right
sheet of the Fermi surface is shown in Fig.\ \ref{FigSPECQLA} for the two
branches $l=0$ and $l=1$. Comparing Figs.\
\ref{FigBlochSpec} and \ref{FigSPECQLA}, we clearly see how the excitation
spectrum is modified by the superconducting pairing. The dispersion of the
pairing amplitude $\Delta ^\alpha _\sigma (q_z,l)$ with respect to $q_z$
increases with decreasing field. For large $N$ and some particular values of
the field (for example in the phase $N=8$ at $H=1.7$ T),
$\Delta ^\alpha _\sigma (q_z,l)$ vanishes at $q_z=\pm \pi /2a_z$ which
results in a gapless excitation spectrum. It is rather
surprising that the phase $N=8$ is gapless at $H=1.7 $ T while the phase
$N=10$ is never gapless. This point, which seems to indicate again that the
validity of the QLA is restricted to the last few phases, will be further
discussed in Sec.\ \ref{SubSecES}. Nevertheless, the
QLA clearly shows the general tendency: the system evolves from a quasi-2D
(BCS-like) behavior in very high magnetic field ($\tilde t\ll 1$) towards
a gapless (or at least with a weak minimum excitation energy) behavior at
weaker field.

In the QLA, the zeros of the excitation
spectrum are related to the zeros of the order parameter in real space. This
results from the relation
\begin{equation}
U_{2m+l_0,l_0}=e^{i\phi }{{(-1)^m} \over {\sqrt{\bar V_{l_0,l_0}}}}
\sum _l U_{2l+l_0,l_0} J_{l-m}(\tilde t)
J_{-l_0-l-m}(-\tilde t) \,,
\label{RELA}
\end{equation}
where $\phi =0$ or $\pi $ depending on the value of the magnetic field.
Eq.\ (\ref{RELA}) will be derived in Sec.\ \ref{SecBQLA}. Using
(\ref{RELA}), the order parameter (\ref{po2}) can be rewritten as
\begin{equation}
\Delta _{l_0,N}({\bf r}) = \Delta e^{-i\phi }\sqrt{\bar V_{l_0,l_0}} (-1)^m
e^{il_0Gx} \sum _p \gamma _p e^{-ipNGx} U_{2m-pN+l_0,l_0} \,.
\end{equation}
Comparing the preceding equation with (\ref{AmpliQLA}), we see that
\begin{equation}
\vert \Delta _\sigma ^\alpha (q_z=\pm {{\pi } \over {2a_z}},l) \vert \sim
\vert \Delta _{l_0,N}(x=\mp {{a_x} \over 4},m=l) \vert  \,.
\end{equation}
This shows that a gapless spectrum at $q_z=\pm \pi /2a_z$ in the
$l$th branch implies that the order parameter in real space
vanishes at the points $(x=\mp a_x/4,m=l)$. Such a relation between the
zeros of the excitation spectrum and the nodes of the order parameter
has also been obtained by Dukan and Te\u{s}anovi\'c in their study of
the vortex lattice of isotropic superconductors in the QLA. \cite{Dukan91}

Unlike for thermodynamics quantities, the third direction has a non trivial
effect on the excitation spectrum. For a 3D system, instead of
(\ref{Exci}), we have
\begin{equation}
E^\alpha _{{\bf q},k_y,l,\sigma }=\pm
\sqrt{ (\epsilon _{q_x,l,\sigma }^\alpha +t_y\cos (k_yb))^2
+\vert \Delta ^\alpha _\sigma (q_z,l) \vert ^2 } \,,
\end{equation}
where $k_y$ is the momentum along the $y$ direction ($-\pi /b<k_y\leq \pi
/b$). $\Delta ^\alpha _\sigma  (q_z,l)$ is the pairing amplitude between the
magnetic Bloch states $\phi _{{\bf q},k_y,l,\sigma }$ and
$\phi _{l_0{\bf G}-{\bf q},-k_y,-l_0-l,\overline \sigma }$. It remains
unchanged since the Cooper pairs are formed with states of opposite momenta
in the $y$ direction. The effect of the third direction can be viewed as
a shift of the MBZ which is now centered around $\pm k_F(k_y)$ defined by
\begin{equation}
v(k_F(k_y)-k_F)+t_y\cos (k_yb)=0 \,.
\end{equation}
Each time $\Delta ^\alpha _\sigma  (q_z,l)$ vanishes (for a given Bloch
momentum $q_z$ and a given branch $l$), there is a line of zeros in the
excitation spectrum determined by
\begin{equation}
\epsilon _{q_x,l,\sigma }^\alpha +t_y\cos (k_yb)=0 \,,
\end{equation}
where $q_x$ belongs to the MBZ and $-\pi /b<k_y\leq \pi /b$. When
calculating thermodynamics quantities, one has to sum over the quantum
numbers associated with the excitation spectrum. For a linearized dispersion
law (Eq.\ (\ref{disp})), the $k_y$ dependence of the spectrum disappears when
summing over the momentum $q_x$ and only yields a factor $1/b$. This
property results from the fact that ${\bf q}$ is coupled only with $l_0{\bf
G}-{\bf q}$. Thus, the direction along the field modifies the quasi-particle
excitation spectrum but not the thermodynamics quantities.

\section{Beyond the quantum limit approximation}
\label{SecBQLA}

In the two preceding sections (Secs.\ \ref{SecTL} and \ref{SecOPQLA}), we have
studied the quantum superconducting phases in the QLA.
This approximation allowed us to
obtain a clear and simple physical picture. Most of the quantities of
interest (critical temperature, Green's functions, thermodynamics
quantities...) could be expressed in a rather simple form. Nevertheless,
the QLA suffers from a lot of weak points. It is restricted to
the quantum regime
$\omega _c\gg T$ and strongly underestimates the critical temperature. It
fails to calculate the free energy in the phases $N\geq 4$. Moreover, as
will be shown in this section, it does not correctly describe the entire
excitation spectrum and the supercurrents vanish in this approximation. It
is therefore necessary to extend the analysis of the two preceding sections
to take into account all the pairing channels. In the QLA, we only
considered the pairing amplitudes $\Delta _\sigma ^\alpha ({\bf q},l;l_0{\bf
G}-{\bf q},-l_0-l)$. In this section, we will call these pairing
amplitudes primary gaps. The other pairing amplitudes will be called
secondary gaps. The reason for this designation is that the primary gaps open
at the Fermi level while the secondary gaps open above and below the Fermi
level (Sec.\ \ref{SubSecES}). Thus, in this section, we will study how
the secondary gaps modify the results obtained in the QLA. In the following,
we first obtain the exact (at the mean-field level) transition line and
construct a variational order parameter, thus recovering the results
obtained in Ref. \cite{Dupuis94a}. Then, from the Gor'kov equations, we
derive the GL expansion of the free energy, and the excitation spectrum is
obtained from the Bogoliubov-de Gennes (BdG) equations. Finally, we
calculate the current distribution in the ordered phase. As in Sec.\
\ref{SecOPQLA}, the PPB effect is ignored.

\subsection{Transition line and variational order parameter}
\label{SubSecTL}

In order to obtain the exact (at the mean-field level) transition line, we
consider the linearized gap equation in a mixed representation $(q_x,m)$ by
taking the Fourier transform with respect to the $x$ direction:
\begin{equation}
\lambda ^{-1} \Delta (q_x,m)=c \sum _{m'} K(q_x,m,m') \Delta (q_x,m') \,,
\label{EQGL}
\end{equation}
\begin{equation}
K(q_x,m,m')= T\sum _{\omega ,\alpha } \int dx \, e^{-iq_x(x-x')}
G_\uparrow ^\alpha ({\bf r},{\bf r}',\omega )
G_\downarrow ^{\overline \alpha }({\bf r},{\bf r}',-\omega ) \,.
\end{equation}
Writing the real space Green's function $G_\sigma ^\alpha ({\bf r},{\bf
r}',\omega )$ as
\begin{equation}
G_\sigma ^\alpha ({\bf r},{\bf r}',\omega )= \sum _{k_x,l}
\phi _{k_x,l}^\alpha ({\bf r}) \phi _{k_x,l}^\alpha ({\bf r}')^*
G_\sigma ^\alpha (k_x,l,\omega ) \,,
\end{equation}
where $\phi ^\alpha _{k_x,l}$ and $G_\sigma ^\alpha (k_x,l,\omega )$
are given in (\ref{Estates}) and (\ref{GF1}), we obtain
\begin{equation}
c K(q_x,m,m')= (-1)^{m-m'} \sum _l V_{l-2m,l-2m'} \chi (q_x+lG) \,.
\label{kernel}
\end{equation}
The matrix $V_{l,l'}$ is defined by (\ref{MatrixV}) and $\chi (q_x)=\sum
_\alpha \chi ^\alpha (q_x)$ is the pair susceptibility at zero
magnetic field evaluated at total momentum $q_x$ (Eq.\ (\ref{PP0})).
It is clear from (\ref{kernel}) that the highest $T_c$ is
obtained when $q_x$ is a multiple of $G$ due to the appearance in this case
of Cooper logarithmic singularities $\chi (0)$. Since the change $q_x
\rightarrow q_x\pm 2G$ is equivalent to the change $m\rightarrow m\pm 1$,
it is sufficient to consider the two cases $q_x=l_0G$ where $l_0=0,1$.
Comparing (\ref{EQGL},\ref{kernel}) with Eq.\ (18) of
Ref. \cite{Dupuis94a}, we note that $\Delta (q_x=l_0G,m)=\Delta
^{Q=l_0G}_{2m}$. The coefficients $\Delta _{2l}^Q$ were introduced in
Ref. \cite{Dupuis94a} through the Fourier expansion of the order parameter.
Thus, we have rederived the linearized gap equation which was obtained in
the gauge ${\bf A}'(0,0,-Hx)$.

In the QLA, where only logarithmic singularities are retained, the
linearized gap equation reduces to
\begin{equation}
\lambda ^{-1}\Delta ^{\rm QLA}(q_x=l_0G,m)=\chi (0) \sum _{m'} (-1)^{m-m'}
V_{2m+l_0,2m'+l_0} \Delta ^{\rm QLA}(q_x=l_0G,m') \,.
\end{equation}
This matrix equation is solved by the orthogonal transformation
$U$ which was introduced in Sec.\ \ref{SecTL} and which diagonalizes the
matrix $V_{l,l'}$. Obviously, we recover the expression (\ref{TCQLA})
of the critical temperature and
\begin{equation}
\Delta ^{\rm QLA}(q_x=l_0G,m)=(-1)^mU_{2m+l_0,l_0} \,.
\end{equation}
For $q_x=(l_0-2p)G$, the order parameter is
expressed as
\begin{equation}
\Delta ^{\rm QLA}(x,m)=(-1)^{m-p} U_{2m-2p+l_0,l_0} e^{i(l_0-2p)Gx} \,.
\label{po4}
\end{equation}
Comparing this expression with (\ref{po1}), we note that we must have
\begin{equation}
U_{2m+l_0,l_0}={\rm const }\,(-1)^m \sum _l U_{2l+l_0,l_0}
J_{l-m}(\tilde t) J_{-l_0-l-m}(-\tilde t)
\end{equation}
in order for Eqs.\ (\ref{po1}) and (\ref{po4}) to be equivalent. The preceding
equation was verified numerically and the exact value of the constant
is given in (\ref{RELA}).

It is not possible to find an analytical expression for the coefficients
$\Delta (q_x=l_0G,m)$. Thus, one has to solve
Eq.\ (\ref{EQGL}) numerically and then calculate every physical quantity as a
function of the coefficients $\Delta (q_x=l_0G,m)$.

As in Sec.\ \ref{SecOPQLA}, we construct the order parameter by taking a
linear combination of the solutions of the linearized gap equation.
$\Delta (q_x=(l_0-2p)G,m)$ has the form of a strip extended in the $x$
direction and localized in the perpendicular direction around $z=(-l_0/2+p)c$.
Thus, the triangular vortex lattice with periodicity $a_z=Nc$ can be
constructed as in Sec.\ \ref{SecOPQLA}, which leads to
\begin{equation}
\Delta (x,m) = \Delta \sum _p \gamma _p \Delta ^{l_0}_{m-p{N \over 2}}
e^{i(l_0-pN)Gx}\,,
\label{po3}
\end{equation}
where $\Delta ^{l_0}_m\equiv \Delta (q_x=l_0G,m)$ is the normalized solution
of the linearized gap equation (\ref{EQGL}): $\sum _m \vert
\Delta ^{l_0}_m\vert ^2=1$. The amplitude $\Delta $ is chosen real.
The coefficients $\gamma _p$ were introduced in Sec.\ \ref{SecOPQLA}.
The order parameter (\ref{po3}) can also be obtained from Eq.\ (41) of
Ref. \cite{Dupuis94a} by the appropriate gauge transformation and using
$\Delta ^{l_0}_m=\Delta ^{Q=l_0G}_{2m}$. Contrary to the QLA, the order
parameter (\ref{po3}) correctly describes the entire phase diagram, from
the weak field regime ($\omega _c\ll T$) where the superconducting state is
an Abrikosov vortex
lattice up to the very high field regime ($\omega _c\gg t_z$) where
the superconducting state is a Josephson vortex lattice. The above form of
the order parameter, obtained by taking a linear combination of the
solutions of the linearized gap equation, is expected to be a very good
approximation as long as the region of interest in the phase diagram is
close to the transition line.

\subsection{Ginzburg-Landau expansion}
\label{SubSecBQLAGLE}

In order to derive the thermodynamics close to the transition line, we have
to perform a GL expansion of the free energy. The situation is more
complicated than in the QLA, because the supercurrents do not vanish and
should therefore be taken into account. The Gor'kov equations have to be
solved in presence of a non uniform flux density ${\bf B}({\bf r})={\bf
H}+{\bf b}({\bf r})$
where ${\bf b}({\bf r})={\bf \nabla }\times {\bf a}({\bf r})$ is obtained
from $4\pi {\bf j}({\bf r})={\bf \nabla }\times {\bf \nabla }\times {\bf
a}({\bf r})$. ${\bf j}({\bf r})$ is the current distribution in the ordered
phase. As noted by Rasolt and Te\u{s}anovi\'c, \cite{Tesanovic92} a
consistent calculation of the free energy to order $\Delta ^4$ requires the
calculation of the Green's functions of the normal phase to first order in
${\bf a}({\bf r})$. Such a procedure, although possible in principle,
greatly complicates the calculation and will not be followed here.
Fortunately, the situation gets simpler in the phase $N=2$. In this phase, the
supercurrents are of order $t_z^2/\omega _c^2$ (Ref.\ \cite{Dupuis94a}) and
their
contribution to the free energy is of order $t_z^4/\omega _c^4$. Thus, to
first order in $t_z^2/\omega _c^2$, the screening of the external magnetic
field can be ignored in the calculation of the free energy. This
approximation is not justified in the phases $N\geq 4$. We hope
that it will give reliable results, at least not too far from the reentrant
phase. Indeed, it will turn out that this approximation is sufficient to
understand the main features of the superconducting phases. Nevertheless,
we will see that a correct description of the phases $N\geq 4$ would require
a proper treatment of the screening of the field by the supercurrents. We
could try to improve the crude approximation $B=H$ by assuming that the flux
density $B$ is uniform but not necessary equal to the external field $H$.
This would however not lead to any significant improvements of the results.
In particular, in the middle of each phase where the averaged flux density
vanishes, this approximation becomes equivalent to completely neglecting
the screening of the field.

As in Sec.\ \ref{SecOPQLA}, we exploit the translational symmetry of the order
parameter using the magnetic Bloch states $\phi ^\alpha _{{\bf q},l}$
(Eq.\ (\ref{BLOCH})). If we ignore the screening of  the external field, the
GL expansion can be written in the form (\ref{freeFE1}) with
\begin{eqnarray}
\alpha &=& {{\lambda ^{-1}} \over S}  \int d^2{\bf r}\,
\Biggl \vert {{\Delta _\sigma ({\bf r})} \over \Delta } \Biggr \vert ^2
-{T \over S} \sum _{\alpha ,\omega ,1,2}
{{\vert \Delta _\sigma ^\alpha (1,2)/\Delta \vert ^2 } \over
{ (i\omega -\epsilon _{1\sigma }^\alpha )(-i\omega -\epsilon
_{2\overline \sigma }^{\overline \alpha }) }} \,,  \nonumber \\
\beta &=& {T \over S}\sum _{\alpha ,\omega } \sum _{1,2,3,4}
{1 \over {\Delta ^4}}
{{\Delta _\sigma ^\alpha (1,2) \Delta _\sigma ^\alpha (1,3)^*
\Delta _\sigma ^\alpha (4,3) \Delta _\sigma ^\alpha (4,2)^* } \over
{ (i\omega -\epsilon _{1\sigma }^\alpha )
  (-i\omega -\epsilon _{3\overline \sigma }^{\overline \alpha })
  (i\omega -\epsilon _{4\sigma }^\alpha )
  (-i\omega -\epsilon _{2\overline \sigma }^{\overline \alpha })  }}
\,,
\label{GLE4}
\end{eqnarray}
where $\Delta _\uparrow ({\bf r})=-\Delta _\downarrow ({\bf r})$ is the
variational order parameter (\ref{po3}). Again, we use the notation $i\equiv
({\bf q}_i,l_i)$. In (\ref{GLE4}), the sums over ${\bf q}_i$ are restricted
to the MBZ defined in (\ref{MBZ1},\ref{MBZ2}).
The pairing amplitudes $\Delta _\sigma ^\alpha (i,j)$ are defined by
(\ref{Ampli}). Using (\ref{po3}), we obtain
\begin{eqnarray}
\Delta _\sigma ^\alpha (1,2)&\equiv &
\delta _{{\bf q}_1+{\bf q}_2,l_0{\bf G}}
\Delta _\sigma ^\alpha (q_{1z},l_1,l_2) \nonumber \\
&=& \delta _{{\bf q}_1+{\bf q}_2,l_0{\bf G}}
\Delta \sum _p \gamma _p e^{-ipq_{1z}a_z} \sum _m \Delta ^{l_0}_m
J_{l_1+p{N \over 2}-m}(\alpha \tilde t)
J_{l_2-p{N \over 2}-m}(\overline \alpha \tilde t) \,.
\label{AmpliE}
\end{eqnarray}
As in the QLA, the total Bloch momentum of two paired states is equal to
$l_0{\bf G}$. But the pairing is not diagonal any more in the branch index
$l$, i.e. $\phi ^\alpha _{{\bf q},l}$ is not coupled only to $\phi
^{\overline \alpha }
_{l_0{\bf G}-{\bf q},-l_0-l}$. Using the property ($p$ integer)
\begin{equation}
\Delta _\sigma ^\alpha (q_z,l_1+pN,l_2-pN)=
e^{2ipq_za_z} \Delta _\sigma ^\alpha (q_z,l_1,l_2) \,,
\end{equation}
we obtain
\begin{eqnarray}
\alpha &=& {c \over {L_z}} \sum _{q_z} \sum _{l_1=1}^N \sum _{l_2}
\Biggl \vert {{\Delta _\sigma ^\alpha (q_z,l_1,l_2) } \over \Delta } \Biggr
\vert ^2 \Bigl \lbrack \chi ((l_0+l_1+l_2)G,T_c) -\chi ((l_0+l_1+l_2)G,T)
\Bigr  \rbrack   \,, \nonumber  \\
\beta &=& {1 \over {L_z}} \sum _{q_z} \sum _{l_1=1}^N \sum _{l_2,l_3,l_4}
{1 \over {\Delta ^4}}
\Delta _\sigma ^\alpha (q_z,l_1,l_2)
\Delta _\sigma ^\alpha (q_z,l_1,l_3)^*
\nonumber \\ & & \times
\Delta _\sigma ^\alpha (q_z,l_4,l_3)
\Delta _\sigma ^\alpha (q_z,l_4,l_2)^*
K_4(l_1,l_2,l_3,l_4) \,,
\end{eqnarray}
where
\begin{eqnarray}
K_4(l_1,l_2,l_3,l_4)&=&  {T \over {L_x}} \sum _{\omega ,\alpha ,k_x}
{1 \over {
(i\omega -\alpha vk_x-\alpha l_1\omega _c)
(-i\omega -\alpha vk_x+\alpha (l_0+l_3)\omega _c) }} \nonumber \\ & &\times
{1 \over {(i\omega -\alpha vk_x-\alpha l_4\omega _c)
(-i\omega -\alpha vk_x+\alpha (l_0+l_2)\omega _c) }} \,.
\label{K4}
\end{eqnarray}
Unless otherwise specified, the sum over the integers $l_i$ runs from
$-\infty $ to $\infty $. In (\ref{K4}), the sum over $k_x$ runs from
$-\infty $ to $\infty $. $\chi (q_x,T)$ is the pair susceptibility
in zero magnetic field at the
temperature $T$. Writing $\alpha =\alpha '(T-T_c)$ with $\alpha '=\partial
\alpha /\partial T \vert _{T_c}$, we obtain, using (\ref{PP0}) and $\Psi (z)
\simeq \ln (z) -1/2z$ for $\vert z\vert \gg 1$:
\begin{equation}
\alpha '\simeq  {{N(0)} \over {T_c}} {c \over {L_z}} \sum _{q_z} \sum _{l=1}^N
\Biggl \vert {{\Delta _\sigma ^\alpha (q_z,l,-l-l_0)} \over \Delta }
\Biggr \vert ^2   \,.
\end{equation}
We also have
\begin{equation}
\beta \simeq \beta _{\rm BCS}{c \over {L_z}} \sum _{q_z} \sum _{l=1}^N
\Biggl \vert {{\Delta _\sigma ^\alpha (q_z,l,-l-l_0)} \over \Delta }
\Biggr \vert ^4 \,.
\end{equation}
The corrections to the above expressions of $\alpha '$ and $\beta $ are of
the order of $T_c^2/\omega _c^2$ (up to logarithmic corrections for $\beta
$). In the quantum regime $\omega _c\gg T$, it is therefore sufficient to take
into account only the primary gaps in order to calculate the coefficients
$\alpha '$ and $\beta $. The secondary gaps, which open at $n\omega _c/2$
($n\neq 0$) above and below the Fermi level (see Sec.\ \ref{SubSecES}), do
not
contribute to $\alpha '$ and $\beta $ to leading order in $T_c/\omega _c$.
However, they have to be taken into account in the calculation of the
critical temperature $T_c$. Comparing the primary gaps in
Figs.\ \ref{FigPAN2}-\ref{FigPAN8}, obtained numerically
from (\ref{AmpliE}), with those obtained analytically in the QLA
(Eq.\ (\ref{AmpliQLA}) and Fig.\ \ref{FigPAQLA}),
it turns out that the result obtained in the QLA is
very close to the exact result. This means that the expression of the
coefficients $\alpha '$ and $\beta $ given by (\ref{GLE2}) is very
accurate. Thus, if we calculate the free energy and the specific heat jump
at the transition, we will recover the results of Sec.\ \ref{SecOPQLA}, the
only difference being that the value of $T_c$ is now exact. Again, we find
that the minimum of the free energy in the reentrant phase is minimized for
$N=2$. When the field decreases from its value in the reentrant phase,
the system undergoes a first order transition and the minimum of the free
energy is then obtained for $N=4$. As in the QLA, the best value of $N$
switches to 6 before reaching the next first order phase transition. This
result, which seems to contradict the assumption that the first order phase
transitions are due to commensurability effects between the period of the
order parameter and the crystalline lattice spacing, indicates that the
contribution of the supercurrents to the free energy has to be taken into
account in the phases $N\geq 4$.

\subsection{Excitation spectrum}
\label{SubSecES}

The excitation spectrum can be obtained from the BdG equations.
\cite{deGennes66} Performing the BdG transformation
\begin{equation}
\psi _\sigma ^\alpha ({\bf r})=\sum _n \Bigl (u_n^\alpha ({\bf r})
\gamma ^\alpha _{n,\sigma } -\sigma v_n^\alpha ({\bf r})^*
\gamma ^{\overline \alpha \dagger }_{n,\overline \sigma } \Bigr ) \,,
\end{equation}
and expanding the coefficients $u$ and $v$ in the basis of the magnetic
Bloch states:
\begin{eqnarray}
u^\alpha _n({\bf r}) &=& \sum _i u_{n,i}^\alpha  \phi _i^\alpha ({\bf
r}) \,, \nonumber \\
v^\alpha _n({\bf r}) &=& \sum _i v_{n,i}^\alpha \phi _i^\alpha ({\bf
r})^* \,,
\end{eqnarray}
we obtain the following BdG equations:
\begin{eqnarray}
(E_{n,\sigma }^\alpha -\epsilon ^\alpha _{{\bf q},l,\sigma })u^\alpha
_{n,{\bf q},l} -\sum _{l'}
v^{\overline \alpha }_{n,l_0{\bf G}-{\bf q},l'}
\Delta _\uparrow ^\alpha (q_z,l,l') &=& 0 \,, \nonumber \\
(E_{n,\sigma }^\alpha +\epsilon ^{\overline \alpha }_{l_0{\bf G}-{\bf
q},l,\overline
\sigma }) v^{\overline \alpha }_{n,l_0{\bf G}-{\bf q},l} -\sum _{l'}
u^\alpha _{n,{\bf
q},l'} \Delta _\uparrow ^\alpha (q_z,l',l)^* &=& 0 \,.
\label{BdGEQ}
\end{eqnarray}
In the quantum regime, $\omega _c\gg T_c$ so that $\omega _c\gg \Delta $
even at $T=0$. The latter inequality allows us to treat the
pairing amplitudes $\Delta ^\alpha _\sigma (q_z,l,l')$ perturbatively.
At low temperature, the order parameter (\ref{po3}) is modified by
contributions of superconducting condensation channels with critical
temperature $<T_c$. However, the main contribution to the order parameter
still comes from the channel with the highest critical temperature so that
(\ref{po3}) should remain a good approximation. To leading order in
$\Delta ^2/\omega _c^2$, the effect of $\Delta ^\alpha _\uparrow (q_z,l,l')$
is to lift the degeneracy between $\epsilon _{{\bf
q},l,\sigma }^\alpha $ and $\epsilon _{l_0{\bf G}-{\bf q},l',\overline
\sigma }^{\overline \alpha }$. Thus, up to corrections of the order
of $\Delta ^2/\omega _c^2$, a gap $2\vert \Delta ^\alpha _\sigma (q_z,l,l')
\vert $ opens in the spectrum at $q_x=\alpha
k_F-(l-l_0-l')G/2$ (with the restriction that $q_x$ has to be in the
MBZ). This gap is located at $\alpha (l_0+l+l')\omega _c/2$ away from the
Fermi level. The primary gaps $\Delta ^\alpha _\sigma (q_z,l,-l_0-l)$
(the only ones which
were considered in the QLA) open at the Fermi level, while the secondary
gaps open above and below the Fermi level. The resulting spectrum, shown
schematically in Fig.\ \ref{FigSPEC}, is
very reminiscent of the one of the FISDW phases.
\cite{Yamaji85,Virosztek86,Montambaux88,Montambaux91}
Since the one-particle states are localized in the $z$ direction on a length
of the order of $c\tilde t$, the pairing amplitudes $\Delta ^\alpha _\sigma
(q_z,l,l')$ are important only if $\vert l-l'\vert <\tilde t$. This means
that there are $\sim N\sim \tilde t$ secondary gaps with a significant value
opening above and below the Fermi level and extending on an energy width of
the order of $t_z$. In the reentrant phase, the secondary gaps are of the
order of $\tilde t^2$ with respect to the primary gaps.

When the field decreases (at $T=0$) below the semiclassical critical field
$H_{c2}(0)$, the amplitude of the order parameter will grow so that $\Delta
$ will become larger than  $\omega _c$. The coherence length $v/\Delta $
then becomes smaller than the (longitudinal) magnetic length $2\pi /G$ and the
quantum effect of the field (i.e., the bending of the semiclassical orbits by
the field) can be ignored. Thus, the condition $\Delta \sim
\omega _c$ (at $T=0$) signals the crossover to the (anisotropic) Abrikosov
vortex lattice  state.

As noted above, the QLA
very accurately describes the primary gaps and therefore provides a
very good approximation of the minimum excitation energy. We also note that
the zeros of the primary gaps which were obtained in the QLA are not
destroyed by the off-diagonal pairings $(l,l')$ ($l'\neq -l_0-l$). A similar
result was obtained in the case of isotropic
superconductors where it has been shown that the gapless behavior obtained
in the QLA is not destroyed by a weak off-diagonal Landau level pairing.
\cite{Dukan91,Norman94}

As in the QLA, we obtain a non monotonous behavior of the minimum excitation
energy: it decreases with the field and vanishes in the phase $N=8$ at $H=1.7$
T, but becomes finite again in the phase $N=10$. This gapless behavior turns
out to strongly depend on the dispersion law of the non-interacting
system and is therefore accidental. For example, if we add a second neighbor
hopping term $t_z'\cos (2k_zc)$ to the dispersion law (\ref{disp}), the
zeros in the excitation spectrum appear for different values of the magnetic
field. A proper treatment of the screening of the magnetic field would
modify the BdG equations and is expected to suppress this accidental gapless
behavior. Nevertheless, when the field decreases, the minimum excitation
energy decreases and becomes very small in the large $N$ phases as can be
seen in Fig.\ \ref{FigPANgd}. This figure shows the gap which opens at the
Fermi level in the phases $N=26$ and $N=28$ (the order parameter $\Delta
_{l_0,N}({\bf r})$ corresponding to the phase $N=26$ is shown in Fig.\ 7 of
Ref.\
\cite{Dupuis94a}). Even if we believe that the gapless behavior of the phase
$N=26$ is accidental, Fig.\ \ref{FigPANgd} clearly shows that the large $N$
superconducting phases effectively become gapless (note that one can
distinguish 21 different phases (i.e. $N=2,\cdot \cdot \cdot ,42$) in the
quantum regime: see Fig.\ 1b of Ref.\ \cite{Dupuis94a}). It is however
difficult to
conclude if the system evolves towards a real gapless behavior. Let
us finally mention that the wrong estimation of the minimum excitation
energy in the phases $N\geq 4$ is related to the poor estimation of the free
energy in these phases.

\subsection{Current distribution}
\label{SubSecCD}

In this section, we give the expression of the current distribution in the
superconducting phases. We will show that the primary gaps do not
contribute to the supercurrents. Since the calculation
is very fastidious, we only give the final expressions.
The current distribution is obtained from
\begin{eqnarray}
j_x({\bf r}) &=& evT \sum _{\omega ,\alpha ,\sigma }
\alpha \delta G_\sigma ^\alpha ({\bf r},{\bf r},\omega ) \,,
\nonumber \\
j_z(x,m,m+1)= &=& - {{ect_z} \over {2i}} (e^{c\partial _z}-e^{c\partial
_{z'}}) T \sum _{\omega ,\alpha ,\sigma } \delta G_\sigma ^\alpha ({\bf
r},{\bf r}',\omega ) \vert _{{\bf r}={\bf r}'} \,.
\label{cur1}
\end{eqnarray}
Here, $j_z(x,m,m+1)$ is the current at point $x$ between the chains $m$ and
$m+1$. $\delta G_\sigma ^\alpha $ is the correction to the Green's function
which results from a non zero order parameter. To lowest order,
\begin{equation}
\delta G_\sigma ^\alpha ({\bf r},{\bf r}',\omega ) =
-\sum _{1,2,3} \phi _1^\alpha ({\bf r}) \phi _2^\alpha ({\bf r}')^*
\Delta _\sigma ^\alpha (1,3) \Delta _\sigma ^{\overline \alpha }(3,2)^*
G_\sigma ^\alpha (1,\omega ) G_{\overline \sigma }^{\overline \alpha }
(3,-\omega ) G_\sigma ^\alpha (2,\omega ) \,,
\label{DeltaG}
\end{equation}
where $\phi _i^\alpha $ is the magnetic Bloch state defined by (\ref{BLOCH})
and we use the notation $i\equiv ({\bf q}_i,l_i)$. Note that to lowest order
($\Delta ^2$), the screening of the external field can be ignored and the
correction $\delta G_\sigma ^\alpha $ can be calculated with the Green's
functions $G_\sigma ^\alpha (i,\omega )$ of the normal phase in the presence
of a uniform magnetic field $H$. From
(\ref{cur1},\ref{DeltaG}), we obtain the following expressions for the
Fourier transform
of ${\bf j}({\bf r})$ at wave vector ${\bf k}=(p_1NG,p_22\pi /Nc)$:
\begin{eqnarray}
j_x(p_1,p_2) &=& -2e \sum _{l_1=1}^N \sum _{l_2,l_3}
F_{l_0+l_1+l_3,l_0+l_2+l_3}
e^{ -i{\pi \over 2} (l_1-l_2-p_1N) -ip_2{\pi \over N}(l_1+l_2-p_1N) }
\nonumber \\ & & \times
J_{p_1N-l_1+l_2} \Biggl (2\tilde t \sin p_2{\pi \over N} \Biggr )
{1\over {L_z}} \sum _{q_z} e^{-ip_1q_za_z}
\Delta _\uparrow ^+(q_z,l_1,l_3) \Delta _\uparrow ^+(q_z,l_2,l_3)^*
\,, \nonumber \\
j_z(p_1,p_2)&=& {{ect_z} \over {iv}} \sum _{l_1=1}^N \sum _{l_2,l_3}
F_{l_0+l_1+l_3,l_0+l_2+l_3} \sum _{\beta =\pm } \beta
e^{-i{\pi \over 2} (l_1-l_2-\beta -p_1N)
-ip_2{\pi \over N}(l_1+l_2-1-p_1N) }
\nonumber \\ & & \times
J_{p_1N-l_1+l_2+\beta } \Biggl (2\tilde t \sin p_2{\pi \over N} \Biggr )
{1\over {L_z}} \sum _{q_z} e^{-ip_1q_za_z}
\Delta _\uparrow ^+(q_z,l_1,l_3) \Delta _\uparrow ^+(q_z,l_2,l_3)^* \,,
\label{cur2}
\end{eqnarray}
where the function $F_{N,M}$ is defined by
\begin{equation}
F_{N,M}=\cases { {1 \over 4\pi ^2T}\,{\rm Im}\,
\Psi '\Bigl ( {1 \over 2}-{{N\omega _c} \over 4i\pi T} \Bigr ) & {\rm
if} $N=M$ \,, \cr {1 \over \pi (N-M)\omega _c}\,{\rm Re}\, \Bigl \lbrack
\Psi \Bigl ({1 \over 2}-{{M\omega _c} \over 4i\pi T} \Bigr ) -
\Psi \Bigl ({1 \over 2}-{{N\omega _c} \over 4i\pi T} \Bigr )
\Bigr \rbrack & {\rm if} $N\neq M$ \,. }
\end{equation}
The function $F_{N,M}$ has been
introduced in Ref. \cite{Dupuis94a} where its expression is given by
Eq.\ (50) for $\omega _c \gg T$.
If we retain only the primary gaps $\Delta _\sigma ^\alpha
(q_z,l,-l-l_0)$, then $F=F_{0,0}=0$ so that the
supercurrents vanish. Thus, in the QLA where the secondary gaps are ignored,
the supercurrents vanish. The current distribution can also be obtained as a
function of the coefficients $\Delta _m^{l_0}$ using the relation
\begin{eqnarray}
\Delta _\uparrow ^+(q_z,l_1,l_2)&=& \Delta \sum _m \Delta _m^{l_0}
{1 \over N} \sum _{p=1}^N \lbrack 1+i(-1)^p \rbrack
e^{i(q_zc-p{\pi \over N})(l_1-l_2)} \nonumber \\
& & \times e^{-i{\pi \over 2}(l_1+l_2-2m)}
J_{l_1+l_2-2m} \Biggl ( 2\tilde t \sin \Biggl \lbrack q_zc-p{\pi \over N}
\Biggr \rbrack \Biggr ) \,.
\end{eqnarray}
Again, we only give the final expressions:
\begin{eqnarray}
j_x(p_1,p_2) &=& -{{4e\Delta ^2} \over {Nc}} \sum _{m_1,m_2} \sum _{l_1,l_2}
\Delta _{m_1}^{l_0} {\Delta _{m_2}^{l_0}}^*
F_{l_0+l_1+2m_1,l_0+l_2+2m_2} e^{-ip_2{\pi \over N}(l_2+2m_2)}
\nonumber \\ & & \times
e^{i{\pi \over 2}(p_1N-2m_1+2m_2)} J_{p_1N-l_1-2m_1+l_2+2m_2}
\Biggl (2\tilde t\sin p_2{\pi \over N} \Biggr )
\nonumber \\ & & \times
\Biggl \langle e^{-iu(p_1N-l_1-2m_1+l_2+2m_2)}
J_{l_1} \Biggl ( -2\tilde t\sin \Biggl \lbrack u+p_2{\pi \over N}
\Biggr \rbrack \Biggr )
J_{l_2}(-2\tilde t \sin u)
\Biggr \rangle \eta _{p_1,p_2} \,,
\label{cur11}
\end{eqnarray}
\begin{eqnarray}
j_z(p_1,p_2) &=& {{2et_z\Delta ^2} \over {ivN}} \sum _{\beta =\pm } \beta
\sum _{m_1,m_2} \sum _{l_1,l_2}
\Delta _{m_1}^{l_0} {\Delta _{m_2}^{l_0}}^*
F_{l_0+l_1+2m_1,l_0+l_2+2m_2} e^{-ip_2{\pi \over N}(l_2+2m_2-1)}
\nonumber \\ & & \times
e^{i{\pi \over 2}(p_1N-2m_1+2m_2+\beta )} J_{p_1N-l_1-2m_1+l_2+2m_2+\beta }
\Biggl (2\tilde t\sin p_2{\pi \over N} \Biggr )
\nonumber \\ & & \times
\Biggl \langle e^{-iu(p_1N-l_1-2m_1+l_2+2m_2)}
J_{l_1} \Biggl ( -2\tilde t\sin \Biggl \lbrack u+p_2{\pi \over N}
\Biggr \rbrack \Biggr )
J_{l_2}(-2\tilde t \sin u)
\Biggr \rangle \eta _{p_1,p_2} \,,
\label{cur22}
\end{eqnarray}
where we use the notation $\langle \cdot \cdot \cdot \rangle =\int _0^{2\pi}
\cdot \cdot \cdot {{du} \over {2\pi }}$. The function $\eta
_{p_1,p_2}$ is defined by
\begin{equation}
\eta _{p_1,p_2}=
\cases { 1 & {\rm if $p_1$ and $p_2$ are even,}
\cr i  & {\rm if $p_1$ and $p_2$ are odd,}
\cr 0 & {\rm otherwise.} }
\end{equation}
Since $\Delta _m^{l_0}=\Delta _{2m}^{Q=l_0G}$, Eqs.\
(\ref{cur11},\ref{cur22}) are analogous to Eqs.\ (48,49) of Ref.\
\cite{Dupuis94a} where the current distribution has been calculated in the
gauge ${\bf A}'(0,0,-Hx)$. The current distribution was calculated
numerically in Ref.\ \cite{Dupuis94a} using (\ref{cur11},\ref{cur22}).
In the reentrant phase, (\ref{cur11},\ref{cur22}) can be simplified by
retaining only the terms of order $\tilde t^2$: the current distribution is
characteristic of a triangular Josephson vortex lattice.
In the phases $N\geq 4$, the current distribution shows a symmetry of a
laminar type and is different from what is obtained in the Abrikosov or
Josephson vortex lattice. \cite{Dupuis94a}

\section{Conclusion}

We have presented a systematic study of the phase diagram of a quasi-1D
superconductor in a high magnetic field. We have obtained the thermodynamics
quantities and the quasi-particle excitation spectrum of the quantum
superconducting phases which are stabilized at high magnetic field as a
result of the magnetic-field-induced confinement of the electrons. The
reentrant phase (very high field limit) is the natural limit for the study
of superconductivity, since the orbital frustration of the order parameter
is suppressed and the screening of the external magnetic field
can be ignored, which considerably simplifies the
analysis. Although we have not taken into account the screening of the field
in the other phases, our analysis clearly shows how the properties of the
system evolves when the field is decreased from its value in the reentrant
phase.

The main results can be summarized as follows. i) In the reentrant phase
($\omega _c\gg t_z$), the
superconducting state is a Josephson vortex lattice. The behavior of the
system is very close to the zero-field BCS situation, up to corrections
of the order of $t_z^2/\omega _c^2$. For instance, the supercurrents are of
the order of $t_z^2/\omega _c^2$ and the specific heat jump at the
superconducting transition $\Delta C/C_N$ is very close to the zero-field BCS
value $(\Delta C/C_N)_{\rm BCS}$. The phase is paramagnetic due to the
positive slope of transition temperature $T_c$ ($dT_c/dH>0$). Gaps open at
the Fermi level on
the whole MBZ. The dispersion of these gaps is of the order of  $t_z^2/\omega
_c^2$ and the minimum excitation energy is finite. Gaps also open at
$n\omega _c/2$ ($n$ integer) away from the Fermi level but are of the
order of $t_z^2/\omega _c^2$ with respect to the gaps opening at the Fermi
level.
ii) In the other phases
($N\geq 4$), the behavior of the system deviates more substantially from the
zero-field BCS situation. The phase is first paramagnetic and then
diamagnetic, which is a consequence of the sign change of the slope of
the transition line $T_c$. The specific heat jump at the transition becomes
smaller than the zero-field BCS value $(\Delta C/C_N)_{\rm BCS}$. There are
$\sim N\sim \tilde t$ gaps with a significant value opening below and above
the Fermi level. The
dispersion  of the gaps which open at the Fermi level increases. The minimum
excitation energy decreases and the quasi-particle spectrum becomes gapless
for large $N$.

Our result have been obtained within a simple model where the electrons
interact through an effective local attractive interaction. Some of these
results would be modified in a more complicated (or realistic) model. For
example, in the case of a $d$-wave (with respect to the $x$ and $y$ axis)
superconductor, the excitation spectrum would be gapless for any value of
the field. The inclusion in our analysis of the PPB effect, which yields to
the formation of a LOFF state, would also result in a gapless
behavior. However, the general structure of the phase diagram does not
depend on the details of the microscopic mechanism which is at the origin of
the attractive electron-electron interaction: the existence of
superconductivity at high magnetic field in a quasi-1D superconductor
originates in the magnetic-field-induced dimensional crossover and does not
rely on a particular model of superconducting pairing.

\section*{ACKNOWLEDGMENTS }
The author would like to thank G. Montambaux for useful discussions on this
work and M. Gabay for interesting discussions on related subjects.
The Laboratoire de Physique des Solides is Unit\'e Associ\'ee au CNRS.

\begin{figure}
\caption{Diagrammatic representation of the ladder approximation for the
two-particle vertex function $\Gamma ^{\alpha \alpha '}({\bf r}_1,{\bf
r}_2;{\bf r}'_1,{\bf r}'_2)$. The zigzag line denotes the attractive
electron-electron interaction. $\alpha $ and $\alpha '$ refer to the sheet
of the Fermi surface. }
\label{FigLA}
\end{figure}

\begin{figure}
\caption{
Solid lines: critical temperature vs magnetic field for $l_0=0$ and $l_0=1$
in the QLA. Dashed lines: exact mean-field critical temperature (see Fig.\ 1
of Ref.\ 19). $T_{c0}=1.5$ K and $t_z=20$ K. }
\label{FigTCQLA}
\end{figure}

\begin{figure}
\caption{ a) Self-energy correction in the Born approximation due to
impurity scattering. b) Vertex correction due to impurity
scattering for the pair propagator. The dashed lines with a cross denote
impurity scattering. }
\label{FigPP}
\end{figure}

\begin{figure}
\caption{ Amplitude and phase of the order parameter $\Delta _{l_0,N}({\bf r})$
(Eq.\ (\ref{po2})) obtained in the QLA in the phase $l_0=0$, $N=2$
($H=5.8$ T). In order to
compare with the exact order parameter obtained in Ref.\ 19, we have made
a gauge transformation from ${\bf A}(Hz,0,0)$ to ${\bf A}'(0,0,-Hx)$. }
\label{FigPON2}
\end{figure}

\begin{figure}
\caption{ As in Fig.\ \ref{FigPON2}, but for the phase $l_0=1$, $N=4$
($H=4$ T). }
\label{FigPON4}
\end{figure}

\begin{figure}
\caption{ As in Fig.\ \ref{FigPON2}, but for the phase $l_0=0$, $N=6$
($H=2.4$ T). }
\label{FigPON6}
\end{figure}

\begin{figure}
\caption{ As in Fig.\ \ref{FigPON2}, but for the phase $l_0=1$, $N=8$
($H=1.7$ T). }
\label{FigPON8}
\end{figure}

\begin{figure}
\caption{ As in Fig.\ \ref{FigPON2}, but for the phase $l_0=0$, $N=10$
($H=1.3$ T). }
\label{FigPON10}
\end{figure}

\begin{figure}
\caption{ Excitation spectrum $\epsilon _{{\bf q},l,\sigma }^\alpha =
\epsilon ^\alpha _{q_x,l,\sigma }$ of the normal phase in the representation
of the magnetic Bloch states $\phi ^\alpha _{{\bf q},l}$. $q_x$ is
restricted to the MBZ (\ref{MBZ1},\ref{MBZ2}).
The figure corresponds to $N=2$. }
\label{FigBlochSpec}
\end{figure}

\begin{figure}
\caption{ Ratio $r=(\Delta C/C_N)/(\Delta C/C_N)_{\rm BCS}$ vs magnetic
field. }
\label{FigSH}
\end{figure}

\begin{figure}
\caption{ Pairing amplitudes $\vert \Delta ^\alpha _\sigma (q_z,l)\vert $ in
the QLA in the phases $N=2$, $N=4$, $N=6$,
$N=8$ and $N=10$. The units are chosen so that ${\rm max}\, \vert \Delta
_\sigma ^\alpha (q_z,l) \vert =1$. }
\label{FigPAQLA}
\end{figure}

\begin{figure}
\caption{ Quasi-particle excitation spectrum $E^+_{{\bf q},l,\sigma }$ in
the QLA in the reentrant phase $l_0=0$, $N=2$ for the branches $l=0$ (a)
and $l=1$ (b). ${\bf q}$ is restricted to the MBZ. The gap opens at the Fermi
level. }
\label{FigSPECQLA}
\end{figure}

\begin{figure}
\caption{ Pairing amplitudes $\vert \Delta _\sigma ^\alpha (q_z,l,L-l-l_0)
\vert $ in the exact mean-field analysis in the phase $N=2$ ($H=5.1$ T).
A gap $2\vert \Delta _\sigma ^\alpha (q_z,l,L-l-l_0) \vert $ opens in the
branch $l$ at $\alpha L\omega _c/2$ away from the Fermi level. The units are
chosen so that ${\rm max}\, \vert \Delta _\sigma ^\alpha (q_z,l,L-l-l_0)
\vert =1$. }
\label{FigPAN2}
\end{figure}

\begin{figure}
\caption{ As in Fig.\ \ref{FigPAN2}, but for the phase $N=4$ ($H=2.8$ T). }
\label{FigPAN4}
\end{figure}

\begin{figure}
\caption{ As in Fig.\ \ref{FigPAN2}, but for the phase $N=6$ ($H=2$ T). }
\label{FigPAN6}
\end{figure}

\begin{figure}
\caption{ As in Fig.\ \ref{FigPAN2}, but for the phase $N=8$ ($H=1.5$ T). }
\label{FigPAN8}
\end{figure}

\begin{figure}
\caption{ Schematic representation of the quasi-particle excitation
spectrum for the right sheet of the Fermi surface. $q_x$ is restricted to
the MBZ. Gaps open at $n\omega _c/2$ ($n$ integer) away from the Fermi
level. The dispersion with respect to $q_z$ is not shown. }
\label{FigSPEC}
\end{figure}

\begin{figure}
\caption{ Pairing amplitudes (primary gaps) $\vert \Delta _\sigma ^\alpha
(q_z,l,-l-l_0) \vert $ in the exact mean-field analysis in the phases $N=26$
($H=0.48$ T) and $N=28$ ($H=0.45$ T). }
\label{FigPANgd}
\end{figure}


\begin{references}

\bibitem{Abrikosov57} A.A. Abrikosov, Sov. Phys. JETP {\bf 5}, 1174 (1957).

\bibitem{Gorkov59} L.P. Gor'kov, Sov. Phys. JETP {\bf 9}, 1364 (1959).

\bibitem{Parks69} See, for example, {\it Superconductivity}, edited by R.D.
Parks (Dekker, New-York, 1969).

\bibitem{Rajagopal66} A.K. Rajagopal and R. Vasudevan, Phys. Lett. {\bf 23},
539 (1966); L.W. Gruenberg and L. Gunther, Phys. Rev. {\bf 176}, 606 (1968).

\bibitem{Tesanovic89} Z. Te\u{s}anovi\'c, M. Rasolt and L. Xing, Phys.
Rev. Lett. {\bf 63}, 2425 (1989) and Phys. Rev. B {\bf 43},288 (1991).

\bibitem{Tesanovic92}
M. Rasolt and Z. Te\u{s}anovi\'c, Rev. Mod. Phys. {\bf 64}, 709 (1992).

\bibitem{Dukan91} S. Dukan, A.V. Andreev and Z. Te\u{s}anovi\'c, Physica
C{\bf 183}, 355 (1991); S. Dukan and Z. Te\u{s}anovi\'c, Phys. Rev. B {\bf
49}, 13017 (1994).

\bibitem{Maniv92} T. Maniv, A.I. Rom, I.D. Wagner and P. Wyder, Phys.
Rev. B {\bf 46}, 8360 (1992); M.J. Stephen, Phys. Rev. B {\bf 45}, 5481
(1992).

\bibitem{Ryan93} J.C. Ryan and A.K. Rajagopal, Phys. Rev. B {\bf 47}, 8843
(1993); A.K. Rajagopal, in {\it Selected Topics in Superconductivity},
edited by L.C. Gupta and M.S. Multani, Frontiers in Solid State Sciences
Vol.1 (World Scientific, Singapore 1993).

\bibitem{Akera91} H. Akera, A.H. MacDonald, and S.M. Girvin, Phys. Rev.
Lett. {\bf 67}, 2375 (1991); M.R. Norman, H. Akera, and A.H. MacDonald,
Physica C{\bf 196}, 43 (1992).

\bibitem{Norman94} M.R. Norman, A.H. MacDonald, and H. Akera, unpublished
(1994).

\bibitem{Rieck90} C.T. Rieck,
K. Sharnberg and R.A. Klemm, Physica C{\bf 170}, 195 (1990); K.
Sharnberg and C.T. Rieck, Phys. Rev. Lett. {\bf 66}, 841 (1991).

\bibitem{Yakovenko93}
V.M. Yakovenko, Phys. Rev. B {\bf 47}, 8851 (1993).

\bibitem{Graebner76} J.E. Graebner and M. Robbins Phys. Rev. Lett. {\bf 36},
422 (1976).

\bibitem{Onuki92} See, for example, Y. Onuki {\it et al.}, J. Phys. Soc. Jpn
{\bf 61}, 692 (1992); R. Corcoran {\it et al.}, Phys. Rev. Lett. {\bf 72},
701 (1994); F.M. Mueller {\it et al.}, Phys. Rev. Lett. {\bf 68}, 3928
(1992); N. Harrison {\it et al.}, Phys. Rev. B {\bf 50}, 4208 (1994);
C.M. Fowler {\it et al.}, Phys. Rev. Lett. {\bf 68}, 534 (1992); G. Kido
{\it et al.}, J. Phys. Chem. Solids {\bf 53}, 1555 (1992); E.G. Haanappel
{\it et al.}, J. Phys. Chem. Solids {\bf 54}, 1261 (1993); R.G. Goodrich
{\it et al.}, J. Phys. Chem. Solids {\bf 54}, 1251 (1993).

\bibitem{Fulde64} P. Fulde and R.A. Ferrell, Phys. Rev. {\bf 135},
A550 (1964); A.I. Larkin and Yu.N. Ovchinnikov, Sov. Phys. JETP
{\bf 20}, 762 (1965).

\bibitem{Lebed86} A.G. Lebed', JETP Lett. {\bf 44}, 114 (1986);
L.I. Burlachkov, L.P. Gor'kov and A.G. Lebed', EuroPhys. Lett. {\bf 4}, 941
(1987).

\bibitem{Dupuis93} N. Dupuis, G. Montambaux and C.A.R. S\'a de Melo, Phys.
Rev. Lett. {\bf 70}, 2613 (1993).

\bibitem{Dupuis94a}
N. Dupuis and G. Montambaux, Phys. Rev. B. {\bf 49}, 8993 (1994).

\bibitem{Dupuis94b}
N. Dupuis, Phys. Rev. B {\bf 50}, 9607 (1994).

\bibitem{Dupuis94c} N. Dupuis, submitted to Phys. Rev. B (1994).

\bibitem{Efetov83} K.B. Efetov, J. Phys. (Paris) Lett. {\bf 44}, L-369
(1983).

\bibitem{Gorkov84} L.P. Gor'kov and A.G. Lebed', J. Phys. (Paris)
Lett. {\bf 45}, L433 (1984).

\bibitem{Lee94} I.J. Lee, A.P. Hope, M.J. Leone and M.J. Naughton,
unpublished (1994).

\bibitem{Bourbonnais91} C. Bourbonnais and L. Caron, Int. J. Mod. Phys. B
{\bf 5}, 1033 (1991).

\bibitem{Nesting} The competition between superconductivity and
spin-density-wave below $T_{x^1}$ is suppressed by assuming that the nesting
of the dispersion law is sufficiently imperfect in the direction of the
field. See Ref.\ \cite{Dupuis94a} for a more detailed discussion.

\bibitem{Melo} This paramagnetic/diamagnetic crossover as the field is
increased within a given phase has also been proposed by C.A.R. S\'a de Melo
(unpublished).

\bibitem{Yamaji85} K. Yamaji, J. Phys. Soc. Jpn {\bf 54}, 1034 (1985).

\bibitem{Virosztek86} A. Virosztek, L. Chen and K. Maki, Phys. Rev. B {\bf
34}, 3371 (1986).

\bibitem{Montambaux88} G. Montambaux and D. Poilblanc, Phys. Rev. B {\bf
37}, 1913 (1988).

\bibitem{Montambaux91} For a review on the FISDW, see for example G.
Montambaux, Physica Scripta T{\bf 35}, 188 (1991).

\bibitem{Dupuis92} N. Dupuis and G. Montambaux, Phys. Rev. B {\bf 46}, 9603
(1992).

\bibitem{Yakovenko87} V.M. Yakovenko, Sov. Phys. JETP {\bf 66}, 355 (1987).

\bibitem{Wannier60} G.H. Wannier, Phys. Rev. {\bf 117}, 432 (1960);
G.H. Wannier, Rev. Mod. Phys. {\bf 34}, 645 (1965); J. Callaway, Phys. Rev.
{\bf 130}, 549 (1963).

\bibitem{Mendez88} E.E. Mendez, F. Agull\'o-Rueda and J.M. Hong, Phys. Rev.
Lett. {\bf 60}, 2426 (1988); P. Voisin, J. Bleuse, C. Bouche, S. Gaillard,
C. Alibert and A. Regreny, Phys. Lett. {\bf 61}, 1639 (1988).

\bibitem{Heritier84} M. H\'eritier, G. Montambaux and P. Lederer, J. Phys.
{\bf 45}, L943 (1984).

\bibitem{Montambaux87} A similar picture of the quantized nesting mechanism
has been given by G. Montambaux, in {\it Low Dimensional Conductors and
Superconductors}, NATO ASI, Vol. 155, Plenum, New-York (1987).

\bibitem{Maki86} K. Maki, Phys. Rev. B {\bf 33}, 4826 (1986).

\bibitem{Gorkov60} L.P. Gor'kov, Sov. Phys. JETP {\bf 10}, 998 (1960).

\bibitem{Anderson59} P.W. Anderson, J. Phys. Chem. Solids {\bf 11}, 26
(1959).

\bibitem{LebedU} A.G. Lebed', private communication (unpublished).

\bibitem{Aslamazov69} L.G. Aslamazov, Sov. Phys. JETP {\bf 28}, 773 (1969);
S. Takada, Prog. Theor. Phys. {\bf 43}, 27 (1970).

\bibitem{Montambaux89} G. Montambaux {\it et al.}, Phys. Rev. B {\bf 39},
885 (1989).

\bibitem{Fetter71} See, for example, A.L. Fetter and J.D. Walecka, {\it
Quantum Theory of Many Particle Systems}, Chap. 13 (McGraw-Hill, 1971).

\bibitem{Kleiner64} W.H. Kleiner, L.M. Roth, and S.H. Autler,
Phys. Rev. {\bf 133}, A1226 (1964).

\bibitem{Poilblanc88} D. Poilblanc, Ph.D thesis, Universit\'e Paris-Sud,
unpublished (1988).

\bibitem{deGennes66} See, for example, P.G. de Gennes, {\it
Superconductivity of Metals and Alloys} (Addison-Wesley, 1966).


\end{references}
\end{document}